\documentclass{article}

\usepackage{graphicx}
\usepackage{epsfig}
\usepackage{amssymb}
\usepackage{amsmath}
\usepackage{amsfonts}
\usepackage{color} 
\usepackage{cite}
 
\newcommand{\ze}{\kern 0.05em}

\begin{document}
\begin{center}

{\LARGE \bf Critical solutions of %Abelian %and non-Abelian 
scalarized
%hairy 
black holes}
\vspace{0.8cm}
\\
{{\bf Jose Luis Bl\'azquez-Salcedo$^{\flat \Diamond}$,
%Carlos A. R. Herdeiro$^{\ddagger}$,  \\
Sarah Kahlen$^{\Diamond}$, Jutta Kunz$^{\Diamond}$,
%Alexandre M. Pombo$^{\ddagger}$ and 
%Eugen Radu$^{\ddagger}$ 
}
\\
\vspace{0.3cm}
$^{\flat}${\small Departamento de F\'isica Te\'orica II, Facultad de Ciencias F\'isicas, 
Universidad Complutense de Madrid, \\ 28040 Madrid, Spain}
\vspace{0.3cm}
\\
$^{\Diamond}${\small Institut f\"ur  Physik, Universit\"at Oldenburg, Postfach 2503,
D-26111 Oldenburg, Germany}
\vspace{0.3cm}
\\
%$^{\ddagger }${\small Departamento de Matem\'atica da Universidade de Aveiro and } \\ {\small  Centre for Research and Development  in Mathematics and Applications (CIDMA),} \\ {\small    Campus de Santiago, 3810-183 Aveiro, Portugal}
}
\vspace{0.3cm}
\end{center}

\date{\today}

\begin{abstract}
We consider charged black holes with scalar hair obtained in a class of Einstein-Maxwell-scalar models,
where the scalar field is coupled to the Maxwell invariant with a quartic coupling function.
Besides the Reissner-Nordstr{\"o}m black holes, these models allow for black holes with scalar hair.
Scrutinizing the domain of existence of these hairy black holes, we observe a critical behavior:
%for fixed coupling constant and horizon radius, and increasing charge, 
A limiting configuration is
encountered at a critical value of the charge, %$Q_{\rm cr}$, 
where spacetime splits into two parts:
%divides into two regions:
%at a critical radial coordinate $r_{\rm cr}=Q_{\rm cr}$.
%The exterior part $r>r_{\rm cr}$ corresponds to the exterior region of an extremal Reissner-Nordstr{\"o}m black hole,
%whereas the interior part $r<r_{\rm cr}$ carries a finite scalar field, that vanishes at $r_{\rm cr}$.
an inner spacetime with a finite scalar field and an outer extremal Reissner-Nordstr{\"o}m spacetime.
Such a pattern was first observed in the context of gravitating non-Abelian magnetic monopoles 
and their hairy black holes.
\end{abstract}

%\pacs{}

%\tableofcontents

%\newpage

\section{Introduction}

In recent years, studies of black holes with scalar hair received much interest, both in the context of
generalized gravity theories such as Einstein-scalar-Gau\ss -Bonnet (EsGB)  theories
\cite{Doneva:2017bvd,Antoniou:2017acq,Silva:2017uqg,Antoniou:2017hxj,Blazquez-Salcedo:2018jnn,Doneva:2018rou,Minamitsuji:2018xde,Silva:2018qhn,Brihaye:2018grv,Doneva:2019vuh,Myung:2019wvb,Cunha:2019dwb,Macedo:2019sem,Hod:2019pmb,Collodel:2019kkx,Bakopoulos:2020dfg,Blazquez-Salcedo:2020rhf,Blazquez-Salcedo:2020caw,Dima:2020yac,Doneva:2020nbb,Berti:2020kgk,Herdeiro:2020wei},
but also in the context of simpler models such as Einstein-Maxwell-scalar (EMs) models
\cite{Herdeiro:2018wub,Myung:2018vug,Boskovic:2018lkj,Myung:2018jvi,Fernandes:2019rez,Brihaye:2019kvj,Herdeiro:2019oqp,Myung:2019oua,Astefanesei:2019pfq,Konoplya:2019goy,Fernandes:2019kmh,Herdeiro:2019tmb,Zou:2019bpt,Brihaye:2019gla,Astefanesei:2019qsg,Blazquez-Salcedo:2020nhs,Blazquez-Salcedo:2020jee,Astefanesei:2020xvn}.
In both cases, the way of coupling the scalar field to the respective invariants, 
the Gau\ss -Bonnet- and the Maxwell invariant,
is crucial for the resulting types of black hole solutions and their properties.
In  fact, according to the choice of coupling function, distinct classes arise.
In the first class, black holes with scalar hair are present, but no general relativity (GR) black holes.
Examples here are dilatonic black holes, first obtained long ago \cite{Gibbons:1987ps,Kanti:1995vq}.
In contrast, the second class allows for both black holes with scalar hair and GR black holes.
In the latter case, we should distinguish whether the GR black holes can exhibit a tachyonic instability,
causing spontaneous scalarization of black holes \cite{Doneva:2017bvd,Antoniou:2017acq,Silva:2017uqg,Herdeiro:2018wub},
or whether the GR black holes will never succumb to such an instability \cite{Astefanesei:2019pfq}.

An example of the latter has been studied in \cite{Blazquez-Salcedo:2020nhs,Blazquez-Salcedo:2020jee}.
In this set of models, the coupling function 
$f(\Phi)=1+\alpha \Phi^4$ has been chosen, coupling the scalar field $\phi$ to the Maxwell invariant with
coupling constant $\alpha$.
Clearly, this coupling function allows for the GR black hole solutions, since the resulting
coupled set of field equations can always be satisfied with a vanishing scalar field.
Thus, the Reissner-Nordstr{\"o}m (RN) black holes are solutions of this set of models
with their usual domain of existence, limited by the set of extremal RN black holes.
However, this set of models allows also for black holes with scalar hair.
These form two branches, the cold branch and the hot branch, that have been
labeled according to their horizon temperature \cite{Blazquez-Salcedo:2020nhs,Blazquez-Salcedo:2020jee}.
For fixed coupling $\alpha$, their domain of existence can be expressed in terms of their
charge to mass ratio $q=Q/M$. The cold branch resides in the interval $[q_{\rm min}(\alpha), 1]$,
and the hot branch in $[q_{\rm min}(\alpha), q_{\rm max}(\alpha)]$, while the bald RN branch resides in $[0,1]$.

On the other hand, hairy black holes with non-Abelian gauge fields have been studied for a long time
(see e.g.~\cite{Volkov:1998cc,Galtsov:2001myk,Kleihaus:2016rgf} for reviews).
Einstein-Yang-Mills (EYM) and Einstein-Yang-Mills-Higgs (EYMS) theories typically not only allow
for black holes with non-Abelian hair but also for embedded Abelian solutions,
such as RN black holes.
This is thus analogous to the EMs models of the second class. 
Unlike most EMs models though, the EYM and EYMH models also feature globally regular solutions, solitons.
In the SO(3)-EYMH case, these solitons correspond to gravitating magnetic monopoles and dyons,
which can be endowed with a horizon, generating hairy black holes
\cite{Lee:1991vy,Breitenlohner:1991aa,Breitenlohner:1994di,Ridgway:1994sm,Brihaye:1998cm,Hartmann:2000gx,Hartmann:2001ic}.
These non-Abelian solutions do not exist for arbitrary values of the coupling constant or horizon radius
but possess a limited domain of existence. As one of the boundaries is approached,
an interesting limiting behavior is observed:
At a critical radial coordinate $r_{\rm cr}=Q_{\rm cr}$, the spacetime divides into two parts.
In the exterior part $r>r_{\rm cr}$, the scalar field assumes its vacuum expectation value,
while the non-Abelian gauge field vanishes except for an embedded Abelian field,
yielding the exterior region of an extremal RN black hole with degenerate horizon $r_{\rm cr}$.
In the interior part $r<r_{\rm cr}$, non-trivial non-Abelian and scalar fields remain. 
These tend to their respective vacuum values at $r_{\rm cr}$.
As this critical solution is approached, both parts of the spacetime become infinite in extent,
since the metric coefficient of the radial coordinate features the double zero of a degenerate horizon.

Here, we will consider the limiting behavior of the EMs hairy black holes on the cold branch,
as the upper boundary of their domain of existence, $q=1$, is approached.
Since the upper limit indeed has $q=1$ and vanishing horizon temperature,
one may expect that an extremal RN solution is approached.
However, as we will show, this is only part of the truth.
In fact, an analogous critical behavior is encountered as has been known
in the case of non-Abelian solutions for long.
As the critical solution is approached, the spacetime splits into two parts,
an exterior part $r>r_{\rm cr}$ corresponding to the exterior region of an extremal RN black hole,
and an interior part  $r<r_{\rm cr}$ with a finite scalar field that vanishes at $r_{\rm cr}$.

In Section II, we present the {EMs} theory 
and the  equations of motion. We then recall the Ansatz for spherically symmetric
black hole solutions and the expansions at the horizon and at infinity.
The properties of the black hole solutions are recalled in Section III.
Next, we consider the critical behavior for fixed coupling constant $\alpha$,
and then address the  $\alpha$-dependence of the properties of the critical solutions in the same section. 
In Section IV, we show that the excited solutions exhibit an analogous critical behavior. We conclude in Section V.

\section{EMs theory}

We consider EMs theory described by the action
\begin{equation}\label{E21}
	 \mathcal{S}=\int d^4 x \sqrt{-g} 
	 \Big[R-2\partial _\mu \Phi \partial ^\mu \Phi -f (\Phi) F_{\mu \nu} F^{\mu \nu} \Big]\ ,
\end{equation}
with the Ricci scalar $R$,  the real scalar field $\Phi$,  the Maxwell field strength tensor $F_{\mu\nu}$
and the coupling function $f(\Phi)$, for which we assume a quartic dependence,
\begin{equation}
f(\Phi) = 1 +\alpha \Phi^4 \ .
\label{coup}
\end{equation}
Assuming a positive coupling constant $\alpha$,
 $\Phi=0$ is the global minimum of the coupling function.

The Einstein-, Maxwell- and scalar field equations follow from the variational principle and read
\begin{eqnarray}
\label{Eins} 
R_{\mu\nu}-\frac{1}{2}g_{\mu\nu}R=
T_{\mu\nu}^{\Phi}+T_{\mu\nu}^{EM} \ , \\
\nabla_{\mu}(\sqrt{-g} f(\Phi) F^{\mu\nu} )
=0 \ , \label{Mxw}
\\
\frac{1}{\sqrt{-g}}\partial_{\mu}
 (\sqrt{-g}g^{\mu\nu}\partial_{\nu}\Phi )
 = \dot f(\Phi) F_{\mu\nu}F^{\mu\nu} \ ,
\label{Klein}
\end{eqnarray}
with $\dot f(\Phi) = d  f(\Phi) /d \Phi$, 
electromagnetic stress-energy tensor
\begin{eqnarray}
T_{\mu\nu}^{EM}\equiv2 f(\Phi)
\left(F_{\mu\alpha}F_{\nu}^{\,\,\alpha}
-\frac{1}{4}g_{\mu\nu}F^2\right)\ ,
\end{eqnarray}
and scalar stress-energy tensor 
\begin{eqnarray}
T_{\mu\nu}^{\Phi}
\equiv\frac{1}{2}\partial_{\mu}\Phi\partial_{\nu}\Phi
-\frac{1}{2}g_{\mu\nu}
\frac{1}{2}(\partial_\alpha \Phi)^2 \ .
\end{eqnarray}

To study static spherically symmetric black holes, 
we employ the metric Ansatz
\begin{eqnarray}
\label{stab1}
ds^2=-N(r) e^{-2 \delta} dt^2 + %\frac{dr^2}{1-2m(r)/r} +r^2(d\theta^2+\sin^2 \theta d\varphi^2) \ ,
\frac{dr^2}{N(r)} +r^2(d\theta^2+\sin^2 \theta d\varphi^2) \ ,
\end{eqnarray}
with the metric functions $N(r)=1-\frac{2m(r)}{r}$ and $\delta(r)$.
We also define the metric function $g(r)=N(r) e^{-2 \delta(r)}=-g_{tt}(r)$.
To obtain black holes with electric charge and, in the scalarized case, also scalar charge,
we parametrize the gauge potential and the scalar field by
\begin{eqnarray}
\label{stab2}
A_\mu =\big( A_t(r),0,0,0\big)  \ , \ \ \ \Phi&=&\Phi(r) \ .
\end{eqnarray} 

Insertion of the above Ansatz leads to the following set of coupled ordinary differential equations (ODEs):
\begin{eqnarray}
\label{E29}
&&
m'=\frac{r^2N \Phi ^{' \ze\ze 2}}{2} +\frac{Q^2}{2r^2 f(\Phi)} \ , \qquad \delta'+r \Phi ^{' \ze\ze 2}=0 \ , \qquad 
 A_t'  = -\frac{Q e^{-\delta}}{f (\Phi) r^2} \ ,
\\
&&
\Phi ^{''} (r)+\frac{1+N}{rN}\Phi ^{'} 
-\frac{Q^2}{r^3N f (\Phi)  }
\left(
\Phi ^{'}
-\frac{\dot{f} (\Phi)}{2r  f (\Phi)}
\right)=0  \ ,
\end{eqnarray}	
where a prime denotes a derivative with respect to the radial coordinate,
and $Q$ is the electric charge of the black holes.

To address the vicinity of the black hole horizon, we perform 
a power series expansion in $r-r_H$ at the horizon,
where we denote the horizon radius by $r_H$
and the horizon values of the functions by the subscript $H$:
			\begin{eqnarray}
&& m(r) = \frac{r_H}{2}+ m_{1} (r-r_H) +\cdots\ \ , \qquad \ \ \ \ \ \ \ \ \ \ \ \ \ \ \ 
			\delta (r)  = \delta _H -\Phi_1^2 r_H (r-r_H)+\cdots\ , \label{nhe}
			\\
&& 
\label{nhe2}
			A_t(r) =\Psi_H -\frac{e^{-\delta _H }Q}{r_H ^2 f (\Phi_H)} (r-r_H)+\cdots
			\ , \qquad \qquad 
			\Phi ( r) =  \Phi _H + \Phi_1 (r-r_H)+\cdots
			\ ,
\end{eqnarray}
with
\begin{equation}
\label{phi1}
	m_{1}=\frac{Q^2}{2r_H ^2 f(\Phi_H)} \ , \ \ \ \ 		 \Phi_1=\frac{Q^2\dot{f} (\Phi_H)}{2 r_H f(\Phi_H) \big[Q^2-r_H^2 {f} (\Phi_H)\big]}~.
\end{equation}			
The global charges of the black holes are read off at spatial infinity, and given by a power series expansion in $1/r$:
	\begin{eqnarray}
	&&
	m(r) =M- \frac{Q^2+Q_s^2}{2r}+\cdots\ ,\qquad  \ \ 
	\delta (r) = \frac{Q_s^2}{2r^2}+\cdots\ ,
	\\
	&&
	A_t(r) =-\frac{Q}{r}+\cdots \ ,\qquad \qquad \qquad \quad 
	\Phi (r) = \frac{Q_s}{r}+\frac{MQ_s}{r^2}+\cdots
	\ ,
	\end{eqnarray}
with ADM mass $M$ and scalar charge $Q_s$.

\section{Limit of cold black holes}

\subsection{Branches of black holes}

We now briefly recall the properties of static spherically symmetric electrically charged
black hole solutions with quartic coupling function (\ref{coup}).
The black holes of the RN branch are given by
\begin{eqnarray}
%N=1-\frac{2M}{r}+\frac{Q^2}{r^2}  \ , \
\delta(r)=0 \ , \
m(r)=M-\frac{Q^2}{2r} \ , \
A_t(r)= -\frac{Q}{r} \ , \
\Phi(r) = 0 \ .
\label{RN_sol}
\end{eqnarray} 

The scalarized black hole solutions are obtained numerically \cite{Blazquez-Salcedo:2020nhs}.
We solve the field equations subject to the boundary conditions that follow from the above expansions
at the horizon and at infinity, with input parameters $\alpha$, $r_H$, and $Q$.
We employ the professional solver COLSYS \cite{Ascher:1979iha},
which is based on a collocation method for boundary-value ODEs and on 
a damped Newton method of quasi-linearization. 
%The problem is linearized and solved at each iteration step, 
%employing a spline collocation at Gaussian points. 
Since this solver includes an adaptive mesh selection procedure, 
it is very suitable for the problem at hand, where high accuracy is
needed in a very small interval close to the black hole horizon.
Consequently, the grid is successively refined until the required accuracy is reached,
typically $10^{-16}$.

The solutions are characterized by a set of dimensionless quantities: 
the charge to mass ratio $q$, the reduced horizon area $a_H$, and the reduced horizon temperature $t_H$, for which 
\begin{equation}
q\equiv \frac{Q}{M} \ , \qquad  a_H\equiv \frac{A_H}{16\pi M ^2} = \frac{ r_H ^2}{4 M ^2}\ ,\qquad t_H\equiv 8\pi M T_H= 2MN'(r_H) e^{-\delta (r_H)}\  \ .
\label{at}
\end{equation}

\begin{figure}[h!]
			 \centering
	 		 \includegraphics[width=0.38\linewidth,angle=-90]{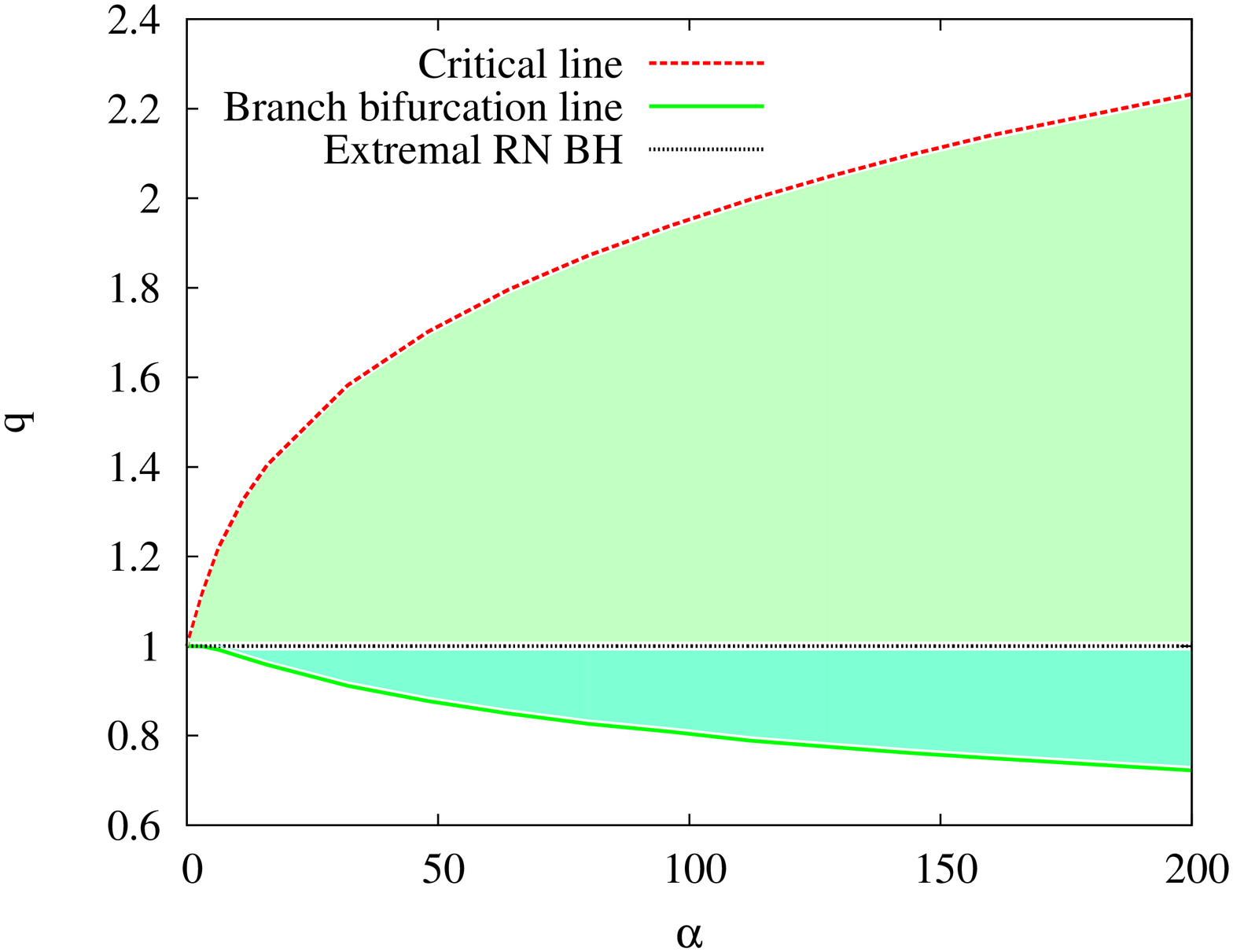}
	 		 \includegraphics[width=0.38\linewidth,angle=-90]{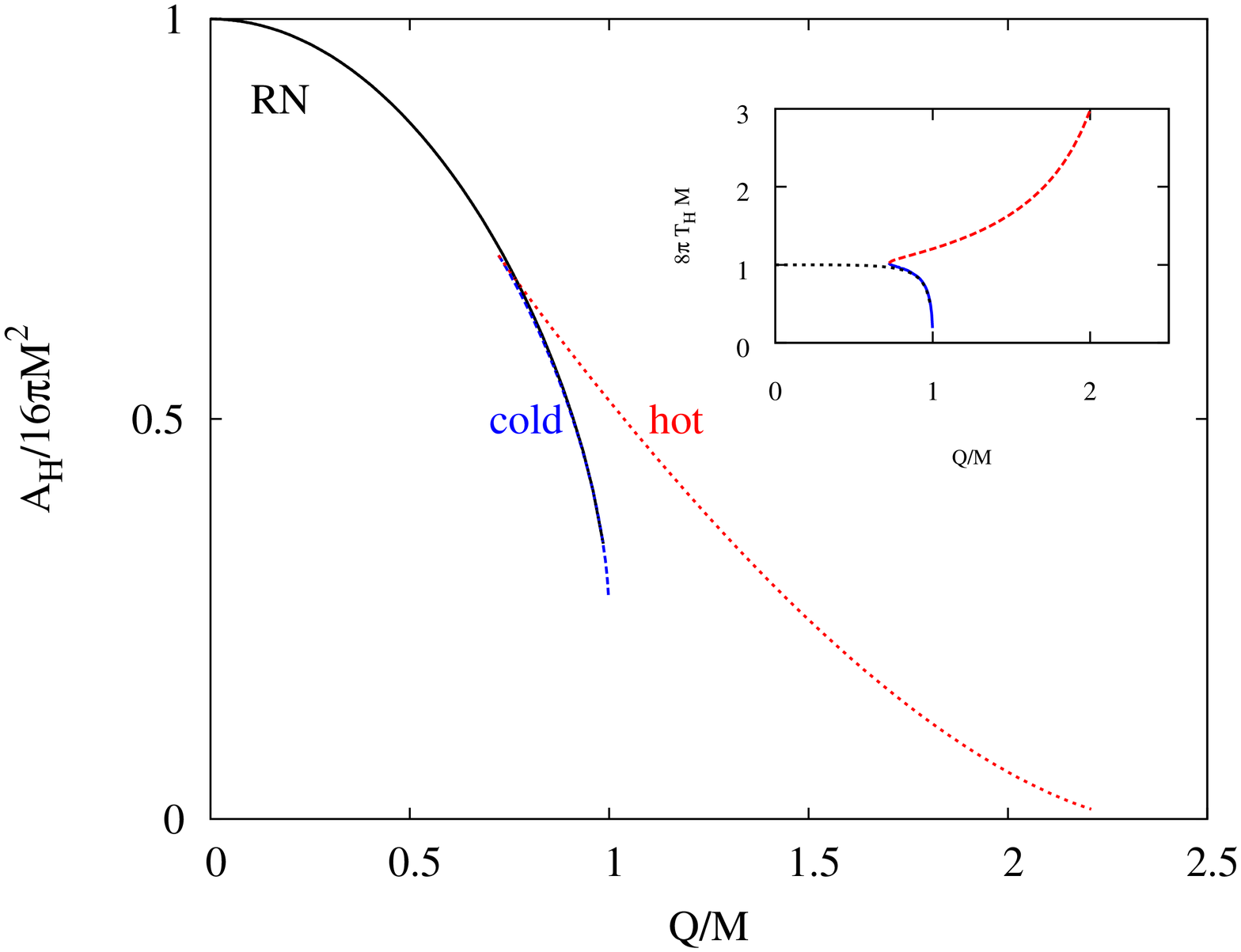}
	 		 \caption{
(a) Phase diagram of scalarized black holes:  mass to charge ratio $q$ vs. coupling constant $\alpha$.
(b) Reduced horizon area $a_H$ (inset: reduced temperature $t_H$) vs. $q$ for $\alpha=200$:
cold branch (blue), hot branch (red) and RN branch (black).
}
\label{fig1}
\end{figure}

We illustrate the domain of existence of the black holes in Fig.~\ref{fig1}(a), where we show
the mass to charge ratio $q$ versus the coupling constant $\alpha$.
The dotted black line $q=1$ represents the set of extremal RN black holes, and thus the upper
boundary for RN black holes.
At the same time, that line represents
the set of critical scalarized solutions forming the upper boundary of the cold black holes,
which reside in the lower green area. The solid green line marks the bifurcation line $q_{\rm min}(\alpha)$
separating the cold and hot black holes. The hot black holes then extend from this bifurcation line
to the dashed red critical line $q_{\rm max}(\alpha)$, i.e., they fill the whole shaded region.

In Fig.~\ref{fig1}(b), we exhibit the reduced area $a_H$ and reduced temperature $t_H$ (inset) versus the
charge to mass ratio $q$ of the black hole solutions for the particular coupling $\alpha=200$ \cite{Blazquez-Salcedo:2020nhs}.
The RN branch is shown in black, the cold branch in blue and the hot branch
in red.  Along the cold branch, the mass to charge ratio $q$ decreases,
while the reduced area $a_H$ and temperature $t_H$ increase.
At the minimal value $q_{\rm min}$, the cold branch bifurcates with the hot branch.
Along the hot branch, $a_H$ decreases again, while $t_H$ increases
with increasing $q$. 

%\section{Limit of cold black holes}

From the figure, it seems that the cold branch
starts from an extremal RN black hole. 
%and then follows the RN branch to a large extent.
Clearly, the charge to mass ratio at its endpoint
agrees with the ratio $q=1$ of an extremal RN black hole,
and the horizon temperature vanishes at its endpoint, $T_H=0$.
%Thus the cold black hole branch seems to reach an extremal black hole configuration.
Looking at the horizon area (Fig.~\ref{fig1}(b)) and further properties,
one is indeed tempted to conclude
that the cold branch emerges from an extremal RN black hole.
However, as we will demonstrate in the following,
this is only partially true. In fact,
the endpoint of the cold branch is %intriguingly different.
a more intriguing configuration.

\subsection{\boldmath $\alpha=200$ \unboldmath}

\begin{figure}[h!]
			 %\centering
			 \mbox{
			 \hspace{-0.8cm}
	 		 \includegraphics[width=0.38\linewidth,angle=-90]{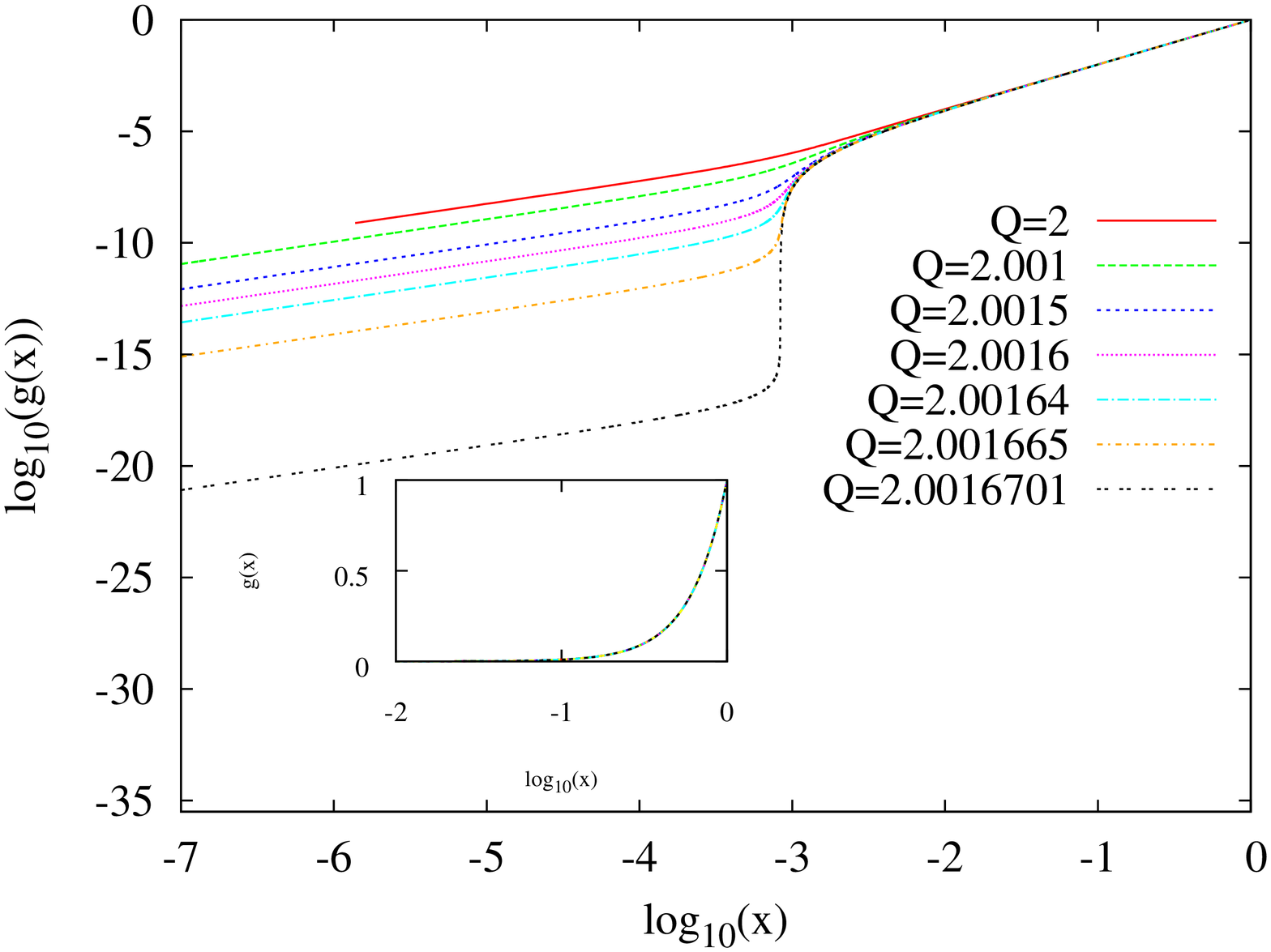}
	 		 \includegraphics[width=0.38\linewidth,angle=-90]{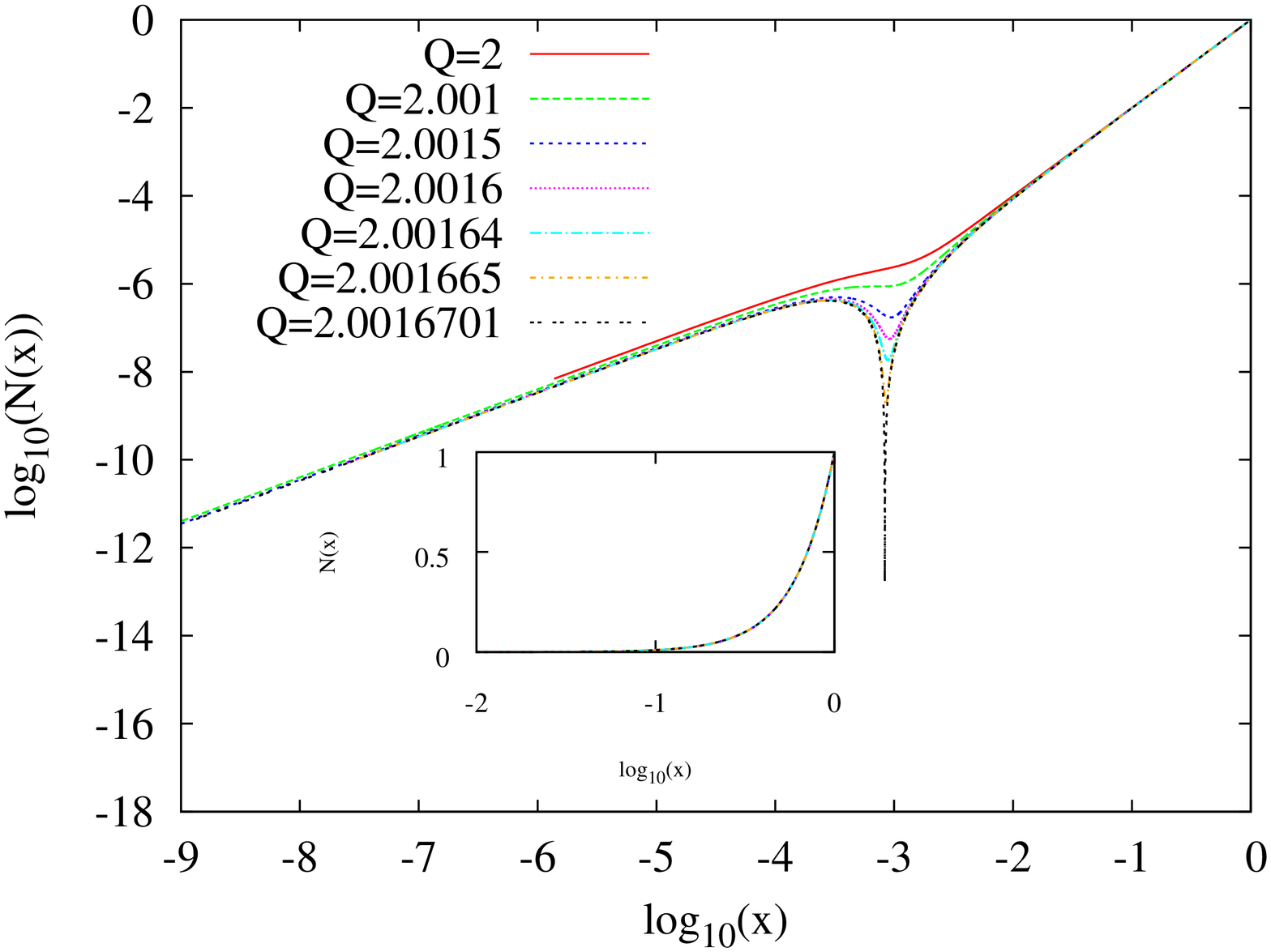}}
	 		 \mbox{
	 		 \hspace{-0.8cm}
	 		 \includegraphics[width=0.38\linewidth,angle=-90]{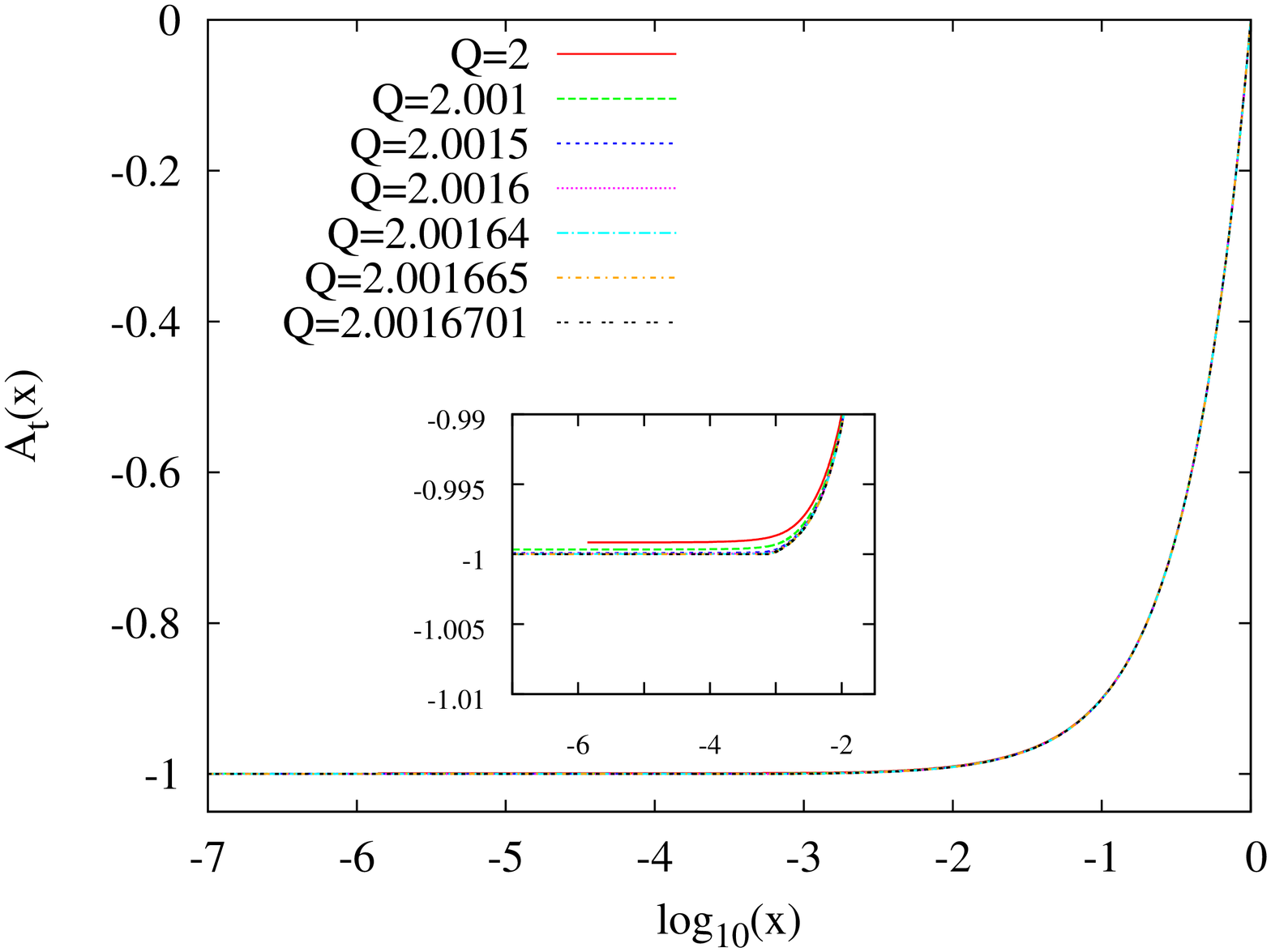}
	 		 \includegraphics[width=0.38\linewidth,angle=-90]{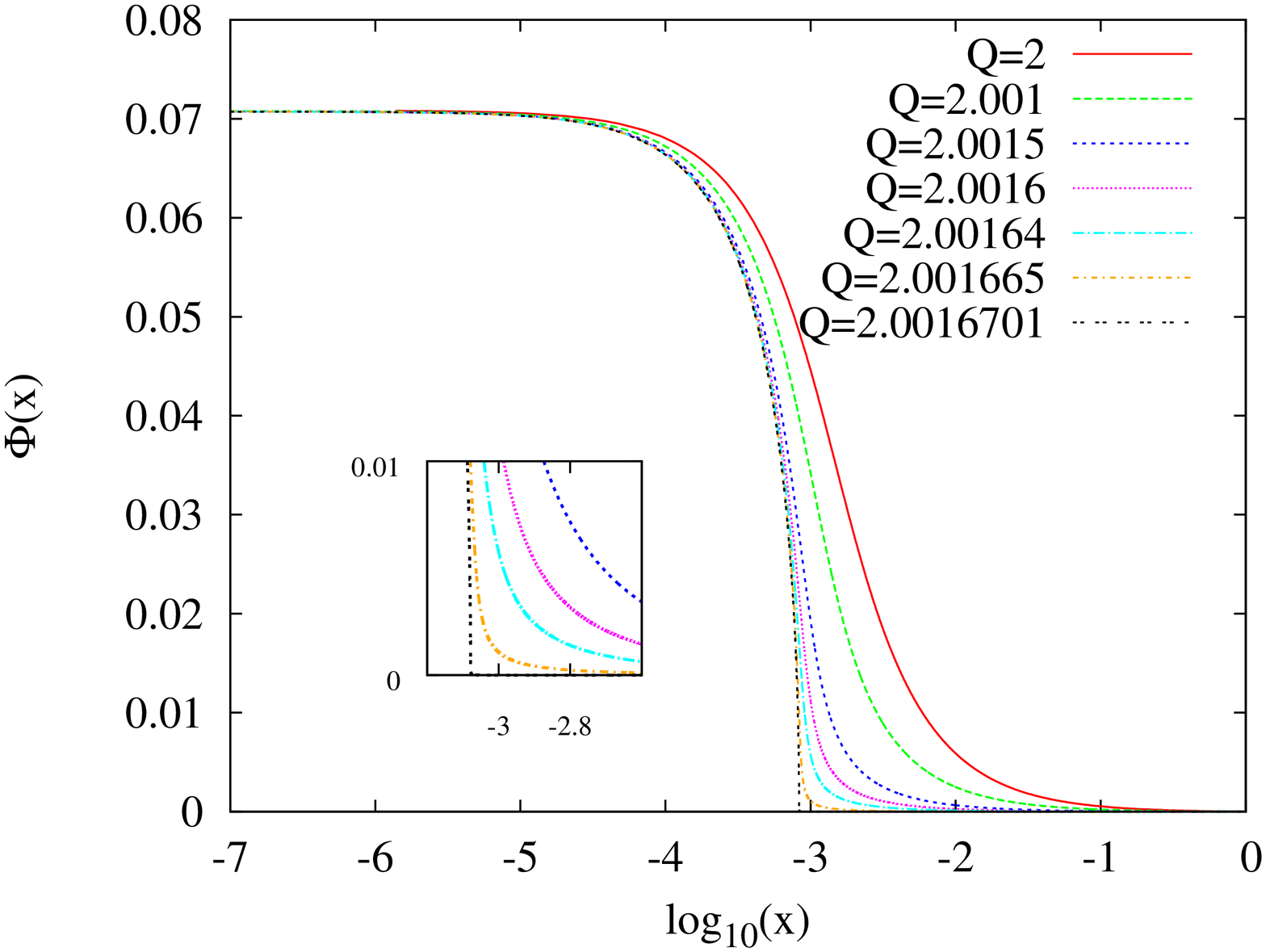}}
	 		 \caption{
Approach to the critical solution for $\alpha=200$:
(a) metric function $g(r)=N(r) e^{-2 \delta(r)}$,
(b) metric function $N(r)=1-\frac{2m(r)}{r}$,
(c) electromagnetic function $A_{t}(r)$
and (d) scalar function $\Phi(r)$
vs. the compactified radial coordinate $x=1-\frac{r_H}{r}$.
The insets highlight the vicinity of the critical radius $r=r_{\rm cr}$.
}
\label{fig2}
\end{figure}

To gain understanding of the limiting configurations,
we now inspect a family of cold black holes with fixed coupling constant $\alpha=200$,
fixed horizon radius $r_H=2$ and increasing charge $Q$.
If the family of cold black holes simply approached an extremal RN black hole,
this extremal black hole would have its degenerate horizon at $r_H=2$,
and it would satisfy $r_H=Q=M$.

In Fig.~\ref{fig2}, we exhibit a set of these cold black hole solutions as they approach 
the endpoint of the cold branch. Notably, all the solutions exhibited here possess a charge
$Q\ge2$, i.e., they range between the extremal RN limit of $Q=2$
and a critical value $Q_{\rm cr}=2.0016701$  ($+ O(10^{-8})$). 
Thus, they slightly exceed the extremal RN limit for fixed horizon radius $r_H=2$.
The figure shows the metric functions $g(r)$ (a) and $N(r)$ (b),
the electromagnetic function $A_t(r)$ (c), and the scalar field function $\Phi(r)$ (d).

Whereas the metric- and electromagnetic functions appear smooth at first glance, as seen in the inlets of Figs.~\ref{fig2}(a) and (b) and in Fig.~\ref{fig2}(c),
the scalar field function (Fig.~\ref{fig2}(d)) immediately reveals a critical behavior:
As $Q_{\rm cr}$ is approached, the scalar field function assumes a 
finite limiting value $\Phi(r_H)$ at the imposed horizon $r_H=2$.
However, the function $\Phi(r)$ decreases more and more steeply
as it approaches zero,  its boundary value at infinity, $\Phi(\infty)=0$.
In fact, in the limit $Q \to Q_{\rm cr}$, it tends to zero already at a critical value of the radial
coordinate, $r=r_{\rm cr}$, which coincides with the value of the critical charge,
$r_{\rm cr}=Q_{\rm cr}$.
In the limit, therefore, the solution features a finite scalar field in the interior $r<r_{\rm cr}$,
whereas the scalar field vanishes identically in the exterior $r>r_{\rm cr}$.

Since the critical exterior solution is a pure electrovacuum solution,
starting at $r_{\rm cr}$ and possessing electric charge $Q_{\rm cr}$,
this suggests that the exterior critical solution is described by an extremal
RN black hole. But instead of carrying charge $Q=2$,
it carries the critical charge $Q_{\rm cr}$.
Comparing the $Q_{\rm cr}$ numerical solution 
with a $Q_{\rm cr}$ extremal RN black hole
shows that this conclusion holds true.
In the interior, however, not only the scalar field function is finite
but all the functions differ from this $Q_{\rm cr}$ extremal RN black hole, as they must
in order to satisfy the imposed boundary conditions at $r_H=2$.

We now inspect the behavior of the functions in the interior in more detail as $Q_{\rm cr}$ is approached, 
starting with the electromagnetic function $A_t$.
As seen in Fig.~\ref{fig2}(c), 
also in the interior, a limiting critical solution is reached.
At the critical radius $r_{\rm cr}$, the electromagnetic function $A_t$ 
of this critical solution assumes the value $A_t(r_{\rm cr})=-1$.
In fact, in the full interior $r<r_{\rm cr}$ it assumes this value,
$A_t(r \le r_{\rm cr})=-1$, as seen in the inset of the figure.
So, there is no electric field in the interior region.

To reveal the critical behavior of the metric functions,
we need to consider double logarithmic plots,
as exhibited in Fig.~\ref{fig2}(a) and Fig.~\ref{fig2}(b).
The extremal RN with charge $Q_{\rm cr}$ would have a double zero
at $r_{\rm cr}$, for both functions $g(x)$ and $N(r)$.
Indeed, we observe a very sharp drop at $r_{\rm cr}$ as the limiting solution is approached for both metric functions
$g(r)$ and $N(r)$, 
confirming our interpretation of the exterior solution.
But in the interior, it becomes apparent that we have not yet
fully reached the critical solution but are only very close to it.

In the interior, both functions $g(r)$ and $N(r)$ differ distinctly.
The function $N(r)$ tends to a finite limiting solution in the interior,
except at $r_H=2$, where the boundary conditions force it to vanish.
In contrast to $N(r)$, the function $g(r)$ approaches zero in the limit.
(With every further digit determined of the critical value $Q_{\rm cr}$,
the function $g(r)$ assumes smaller values in the interior.)
Recalling that the function $g(r)$ has been decomposed into the factors $N(r)$
and $\exp(-2 \delta(r))$, we conclude that it is the function $\delta(r)$ which causes $g(r)$ 
to vanish in the interior in the limit, since $N(r)$ has a finite limit.
It is instructive now to look again at the horizon temperature $T_H$.
%When we observed that $T_H \to 0$ on the cold branch in the limit,
%we might have expected that the reason was that a degenerate horizon would
%arise at $r_H=2$, and therefore the derivative $N'(r)$ would vanish there. But now, we see
%that there is no degenerate horizon at $r_H=2$ in the limit.
Having observed above (e.g. in the inset of Fig.~\ref{fig1}(b)) that $T_H \to 0$ on the cold branch in the limit,
one may have expected that the reason was that a degenerate horizon
arose at $r_H=2$, and therefore the derivative $N'(r)$ vanished there. But now, we see
that there is no degenerate horizon at $r_H=2$ in the limit.
Instead, $T_H$ vanishes because of the factor $\exp(-\delta(r))$ in Eq.~(\ref{at}).

We have thus obtained the following scenario:
In the limit $Q \to Q_{\rm cr}$, the spacetime splits into
an exterior and an interior part.
The exterior is described by an extremal RN black hole;
the interior has a finite scalar field, but no electric field.
Since at the critical radius $r_{\rm cr}$ a double zero
is approached, as featured by a degenerate horizon,
the radial distance $l(r)$ to and from $r_{\rm cr}$
increases as the critical solution is approached.
In the limit, both parts of the spacetime
become infinite in extent.

\begin{figure}[h!]
			 %\centering
			 \mbox{
			 \hspace{-0.8cm}
	 		 \includegraphics[width=0.38\linewidth,angle=-90]{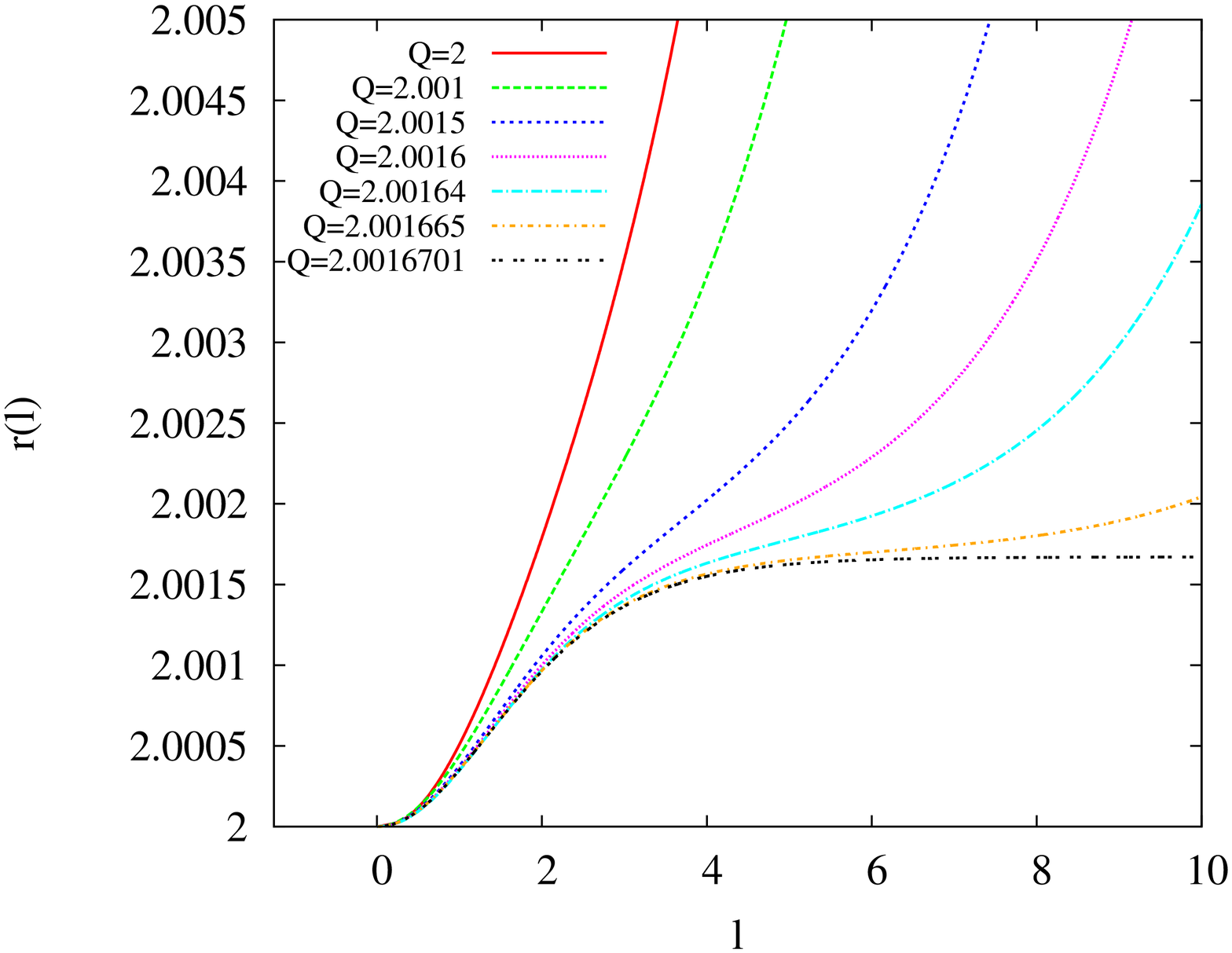}
	 		 \includegraphics[width=0.38\linewidth,angle=-90]{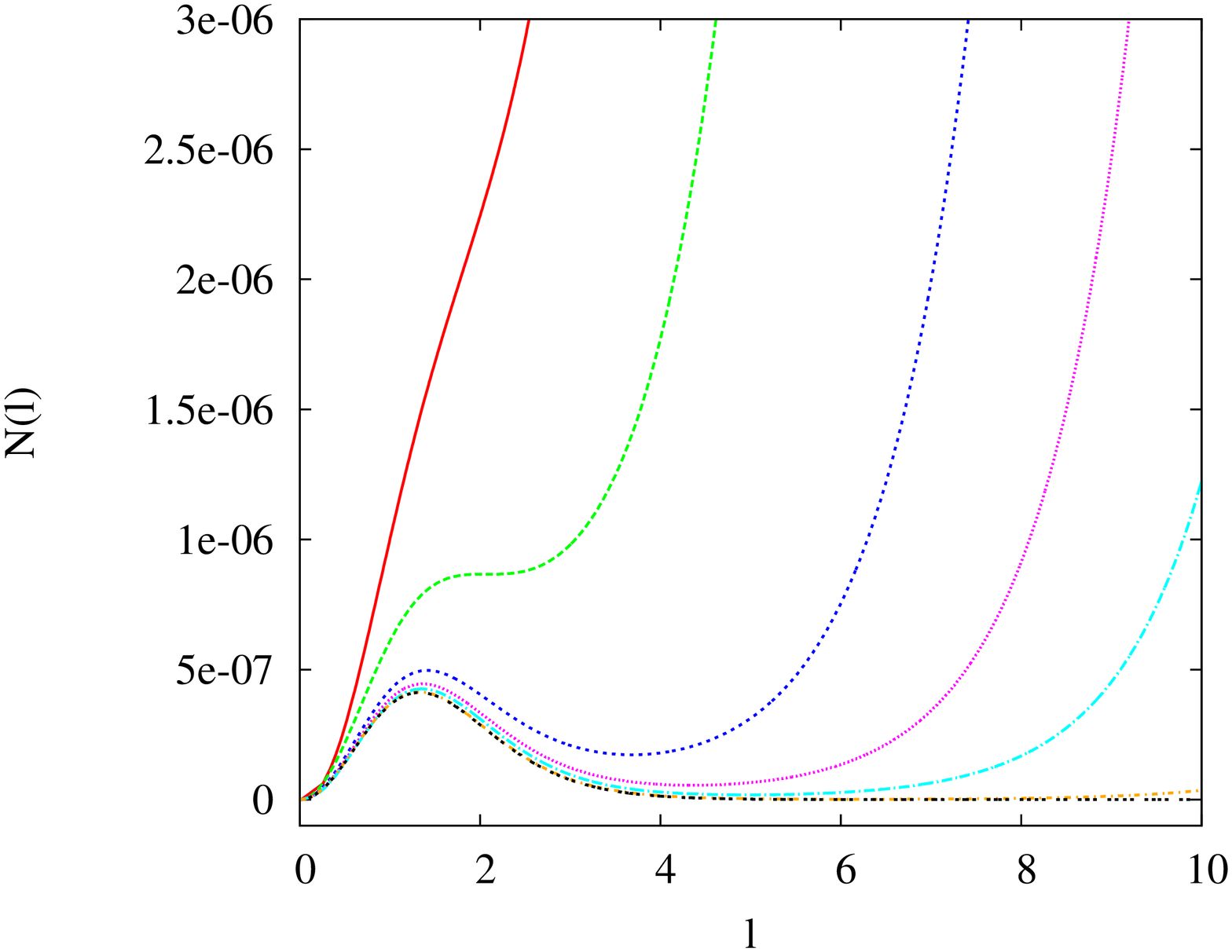}}
	 		 \mbox{
	 		 \hspace{-0.8cm}
	 		 \includegraphics[width=0.38\linewidth,angle=-90]{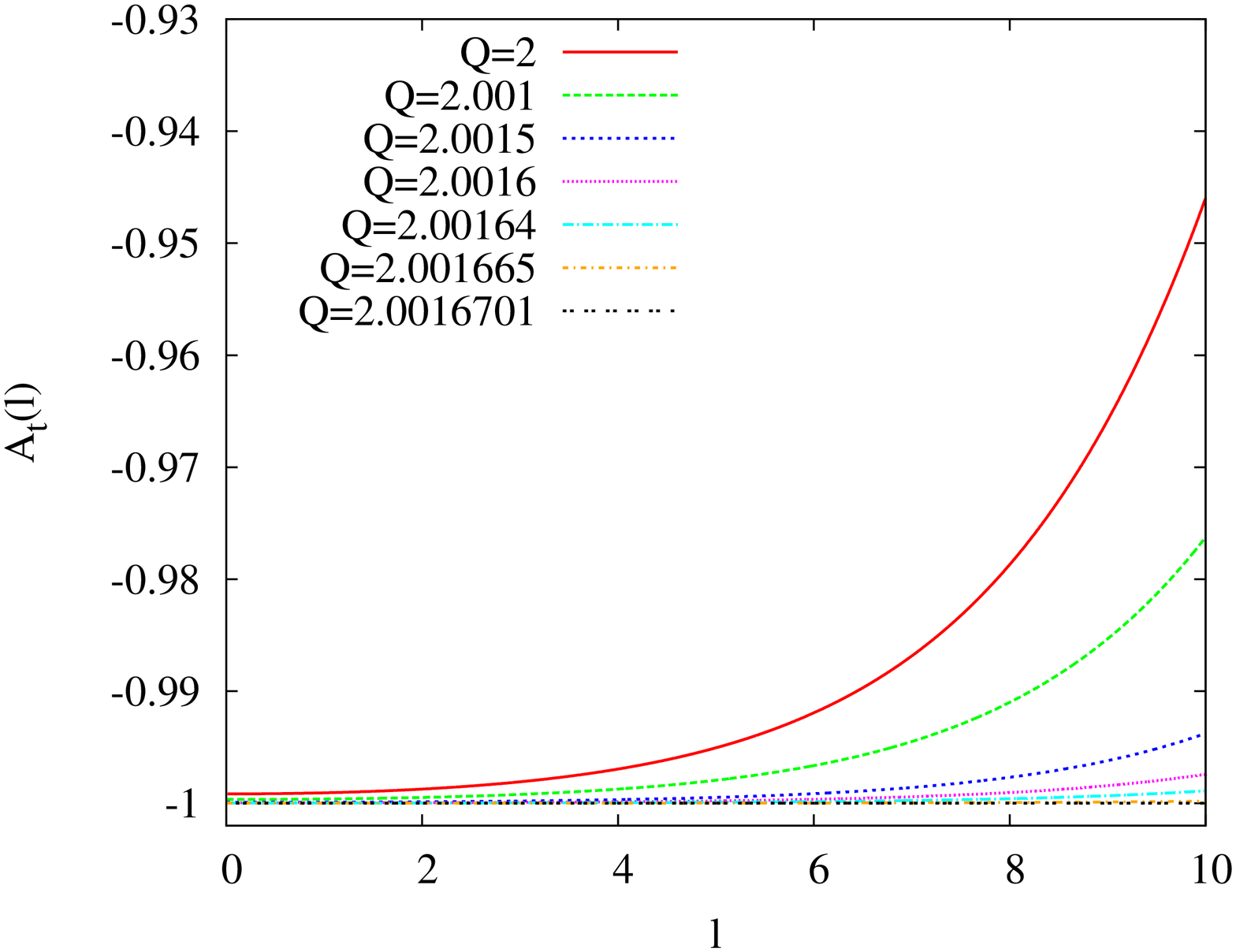}
	 		 \includegraphics[width=0.38\linewidth,angle=-90]{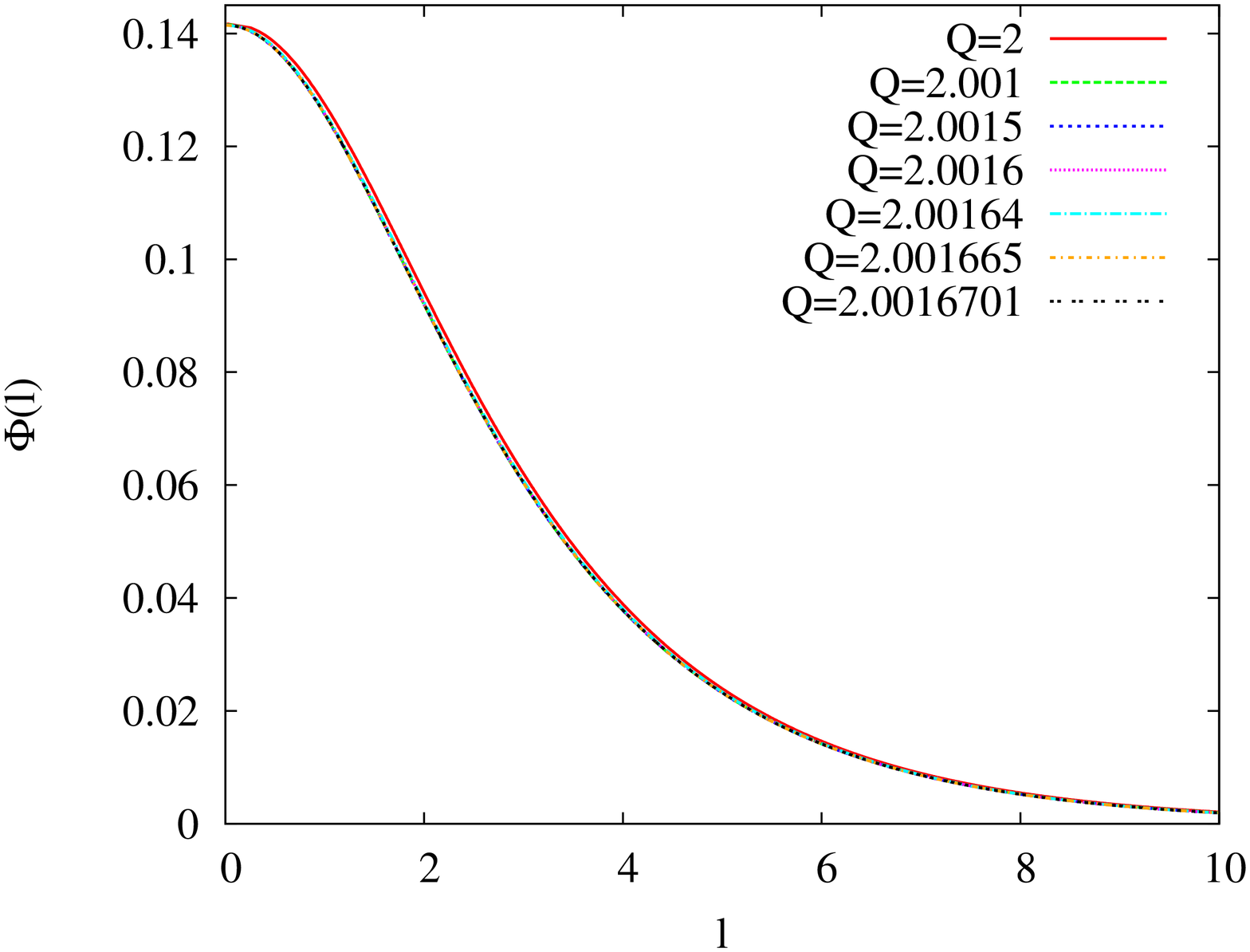}}
	 		 \caption{
Approach to the critical solution for $\alpha=200$:
(a) radial coordinate $r$,
%vs radial distance $l$ (Eq.~(\ref{length})), 
(b) metric function $N(r)=1-\frac{2m(r)}{r}$,
(c) electromagnetic function $A_t(r)$,
and (d) scalar function $\Phi(r)$
vs. the radial distance $l(r)$  (Eq.~(\ref{length})).
}
\label{fig3}
\end{figure}

To demonstrate this effect for the interior part, we consider in Fig.~\ref{fig3} 
the radial distance $l(r)$, defined via
\begin{equation}
l(r) = \int_{r_H}^r \frac{dr}{\sqrt{N(r)}} \  \ .
\label{length}
\end{equation}
The dependence of the radial coordinate $r(l)$ on the radial distance is
illustrated in Fig.~\ref{fig3}(a) for the approach to the critical solution.
For $Q_{\rm cr}$, an infinite radial distance is reached at the finite
value of the radial coordinate $r_{\rm cr}$,
thus an infinite throat is formed.
For $Q \to Q_{\rm cr}$, the metric function $N(r)$ (Fig.~\ref{fig3}(b)),
the electromagnetic function $A_t(r)$ (Fig.~\ref{fig3}(c)),
and the scalar function $\Phi(r)$ (Fig.~\ref{fig3}(d))
are also shown versus the radial distance $l(r)$.
The metric function $N(r)$ again highlights the formation of an infinite
throat in the limit, while the electromagnetic function $A_t(r)$
approaches a constant value in the interior region.
The scalar function $\Phi(r)$, on the other hand, demonstrates
that when considered as a function of the radial distance $l$
instead of the radial coordinate $r$, there is remarkably little
dependence on the value of the charge $Q$ during the approach
$Q \to Q_{\rm cr}$.
An analogous observation was made for the matter functions
of the magnetic monopoles during their approach to their
respective critical solution \cite{Lee:1991vy}.

\subsection{\boldmath $\alpha$-dependence \unboldmath}

\begin{figure}[h!]
			 %\centering
			  \mbox{
			 \hspace{-0.8cm}
	 		 \includegraphics[width=0.38\linewidth,angle=-90]{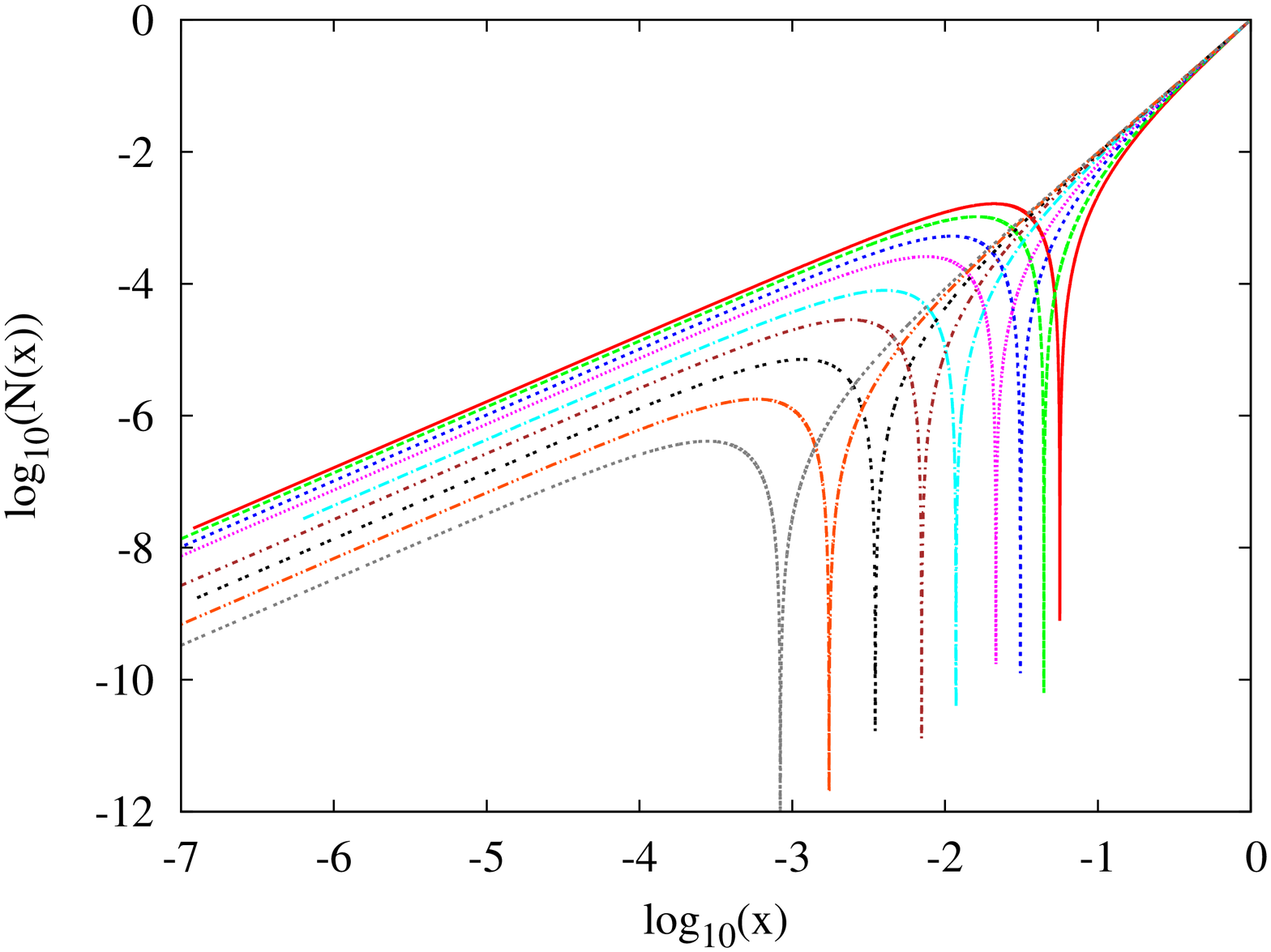}
	 		 \includegraphics[width=0.38\linewidth,angle=-90]{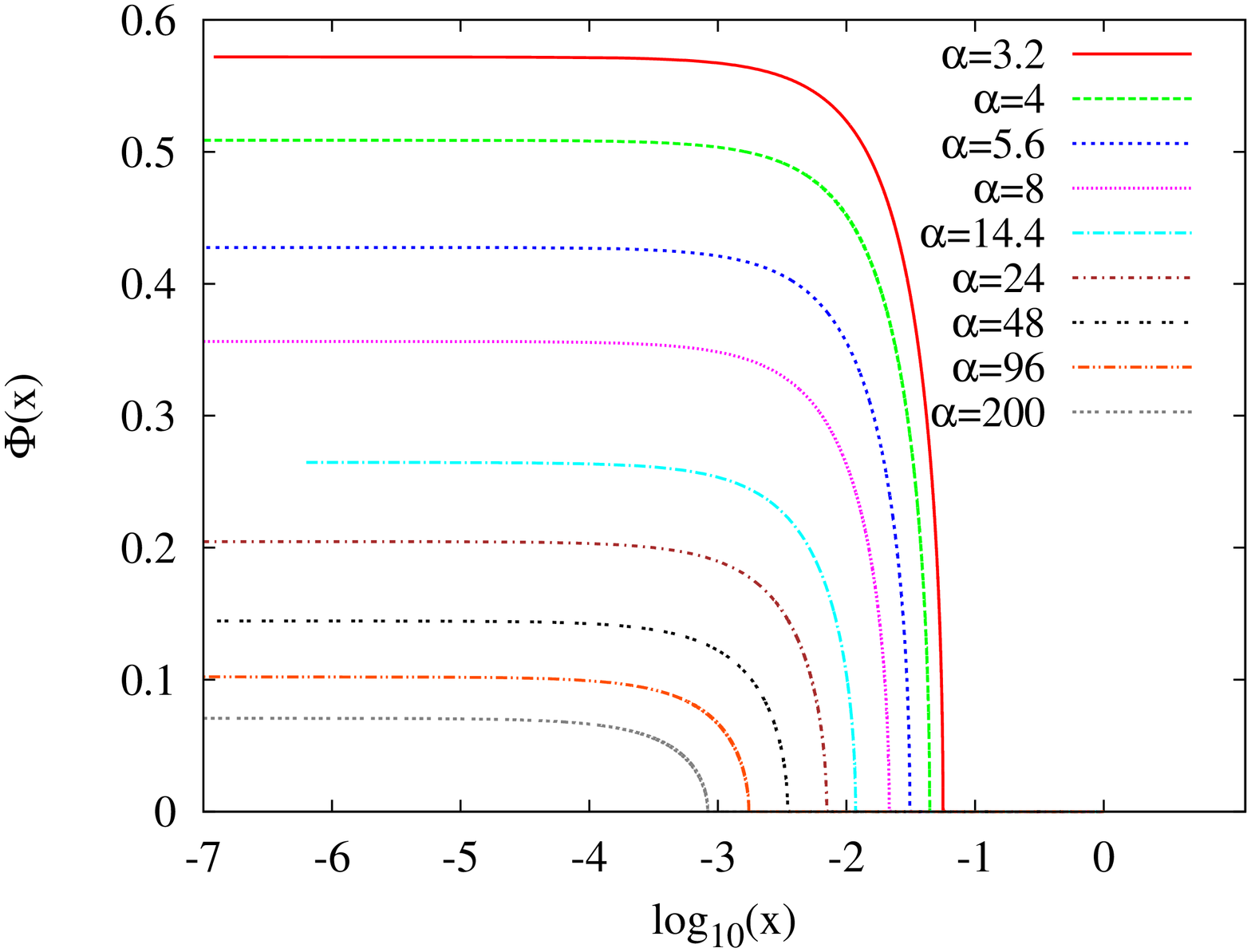}}
	 		 \caption{
	 	Critical solutions for a set of couplings $\alpha$:
(a)  metric function $N(r)=1-\frac{2m(r)}{r}$,
(b)  scalar function $\Phi(r)$
vs. the compactified radial coordinate $x=1-\frac{r_H}{r}$. Note that the key applies to both figures.
}
\label{fig4}
\end{figure}

We now demonstrate that the critical scenario described above 
holds for a large range of couplings $\alpha$, showing that the scenario is rather generic. 
We exhibit the critical solutions in Fig.~\ref{fig4},
where we show the metric function $N(r)=1-\frac{2m(r)}{r}$ (Fig.~\ref{fig4}(a))
and the scalar function $\Phi(r)$ (Fig.~\ref{fig4}(b))
for a set of couplings $\alpha$ in the interval $[3.2,200]$.

\begin{figure}[h!]
			 %\centering
			 \mbox{
			 \hspace{-0.4cm}
	 		 \includegraphics[width=0.38\linewidth,angle=-90]{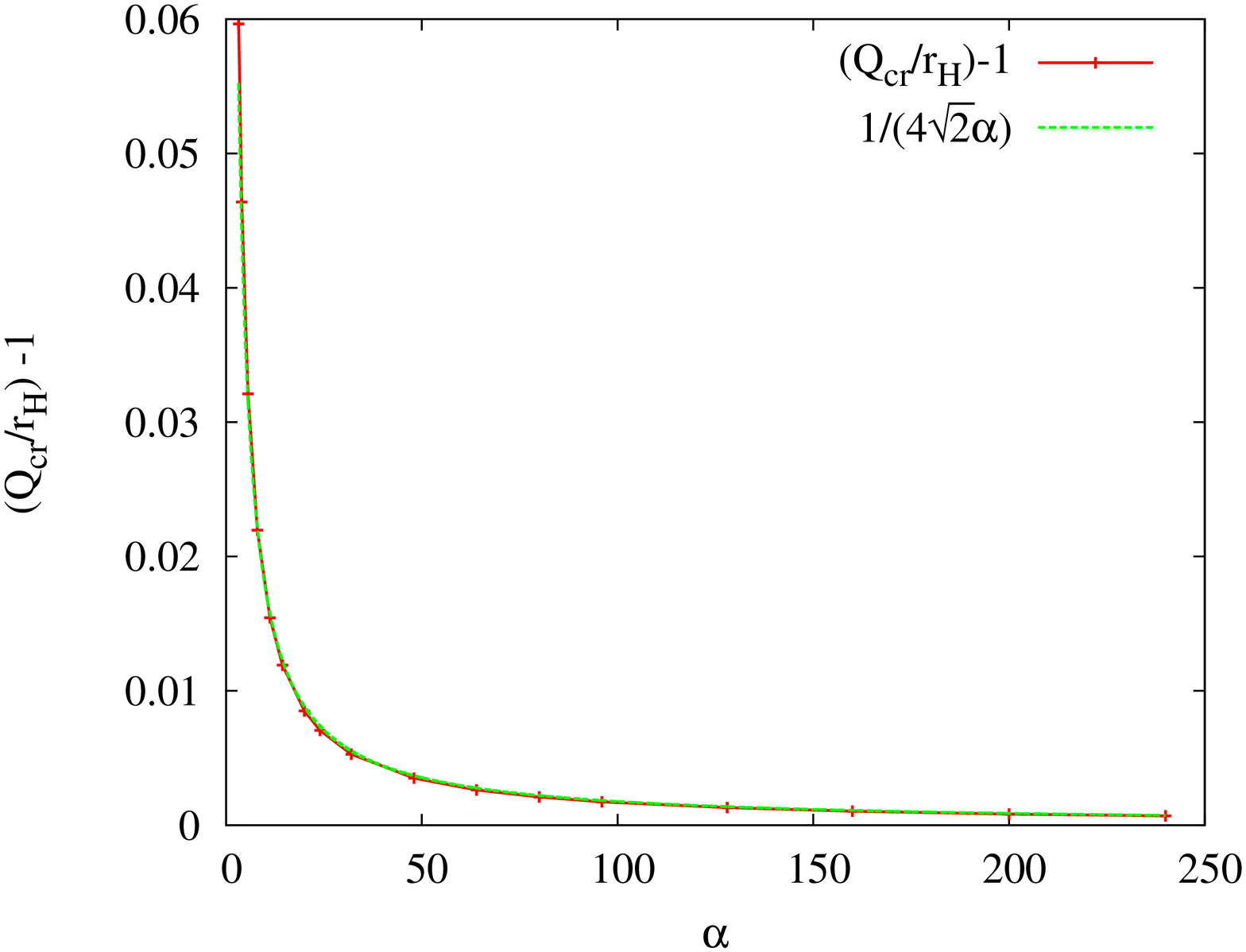}
	 		 \includegraphics[width=0.38\linewidth,angle=-90]{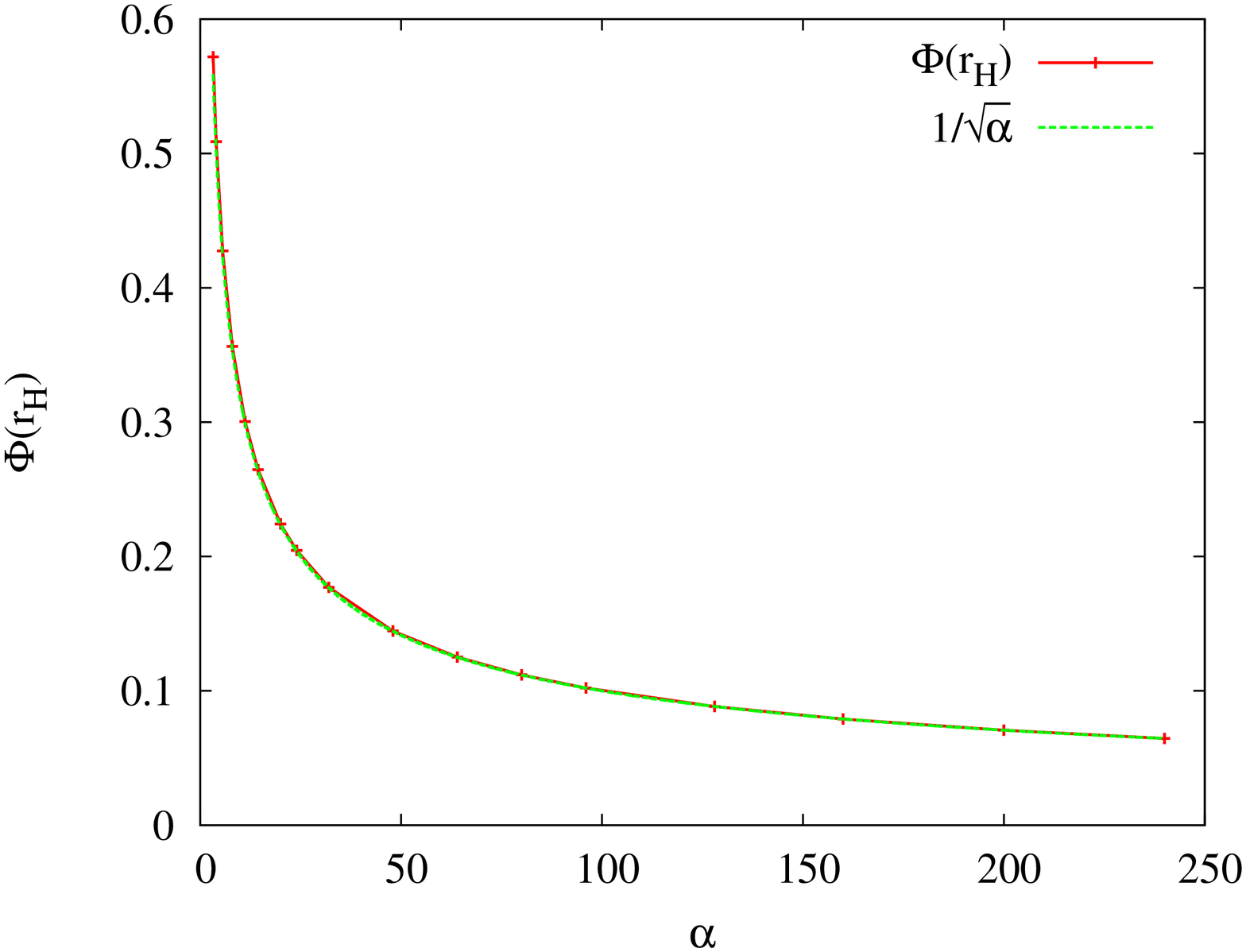}}
	 		 \mbox{
	 		 \hspace{-0.4cm}
	 		 \includegraphics[width=0.38\linewidth,angle=-90]{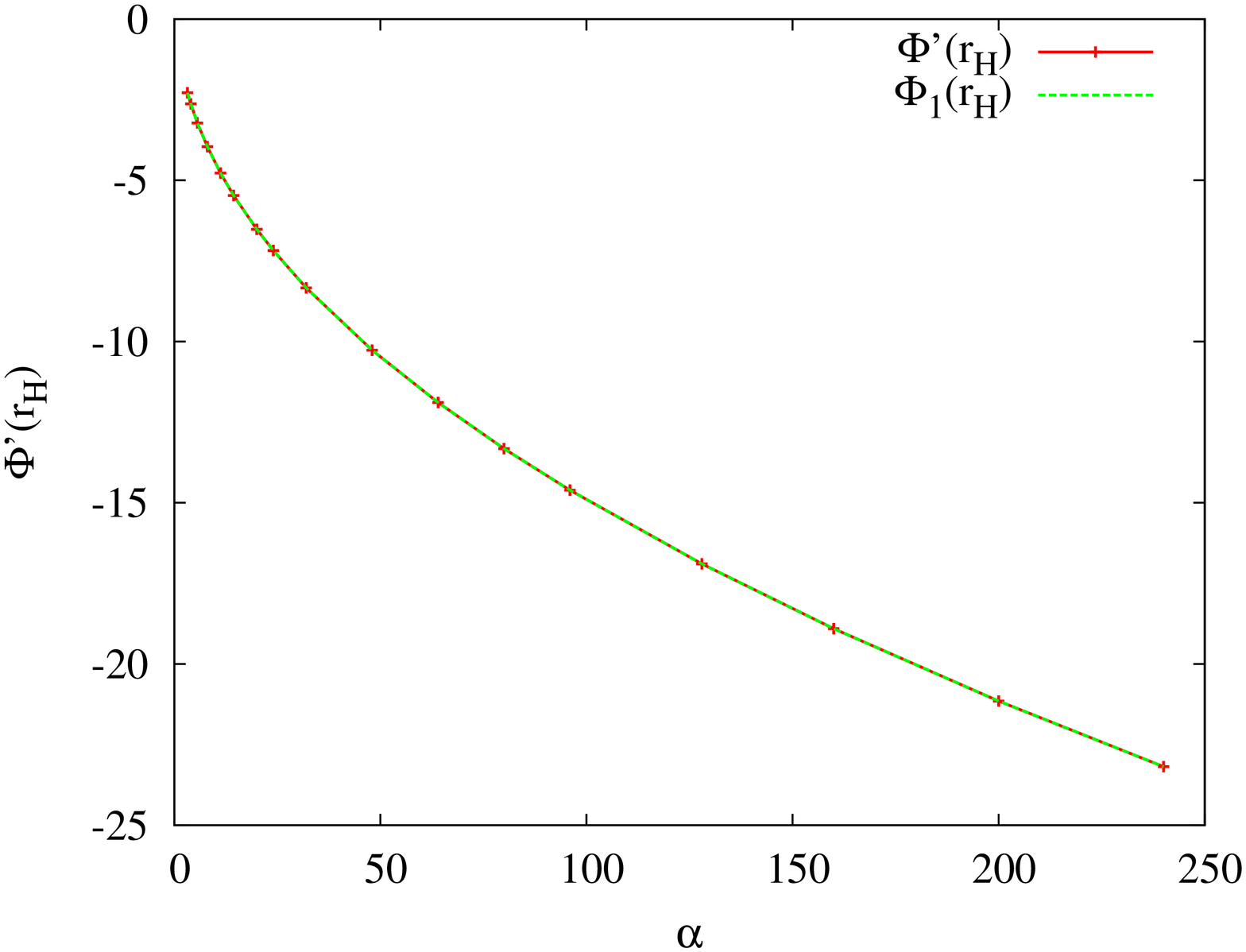}
	 		 \includegraphics[width=0.38\linewidth,angle=-90]{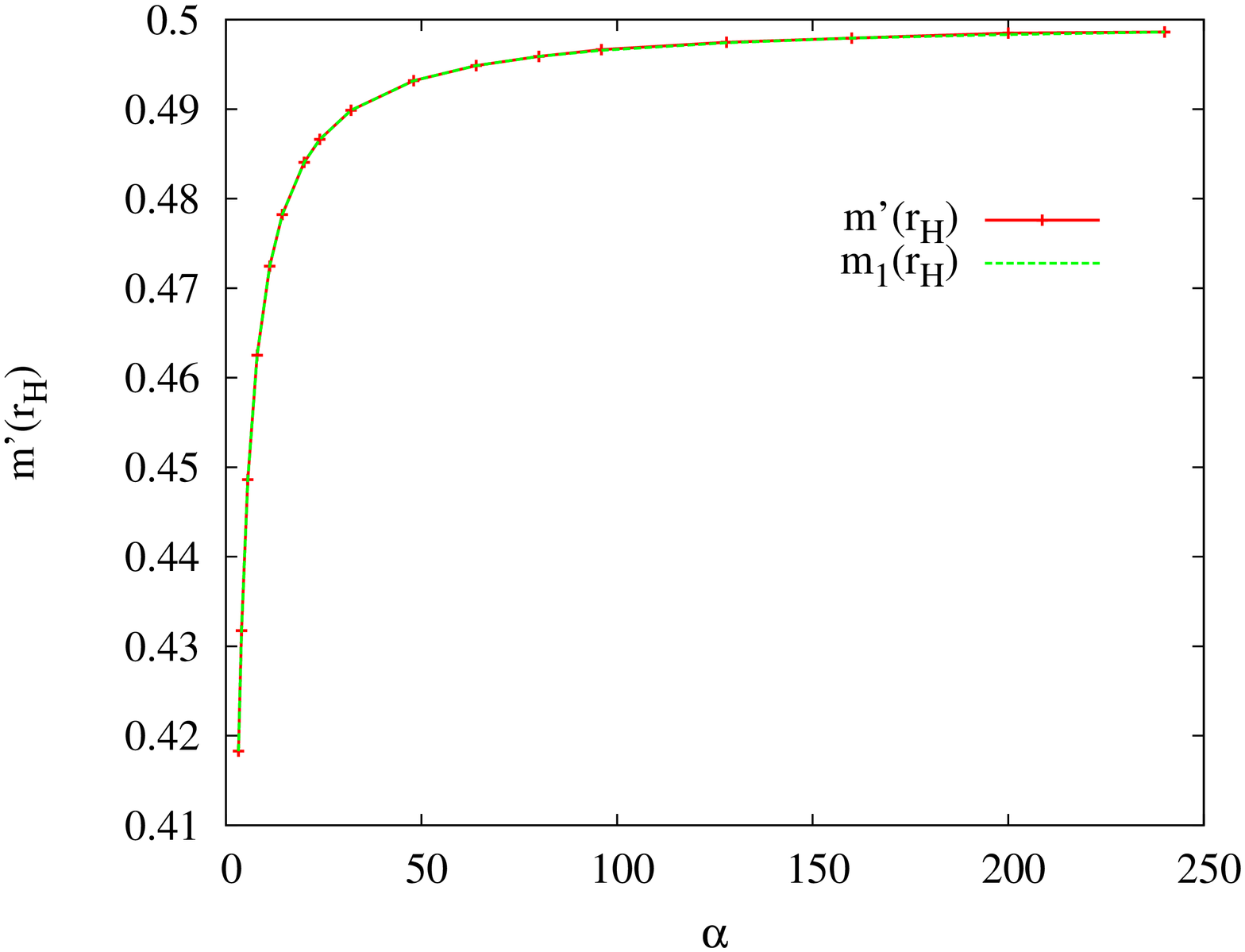}}
	 		% \mbox{
	 		% \hspace{-0.8cm}
	 		% \includegraphics[width=0.33\linewidth,angle=-90]{qcr.eps}
	 		% \includegraphics[width=0.33\linewidth,angle=-90]{phi_h.eps}}
	 		% \mbox{
	 		% \hspace{-0.8cm}linewidth,angle=-90]{mp.eps}}
	 		 \caption{
	 		 Properties of critical solutions:
	 		 (a) $Q_{\rm cr}(\alpha)$ (red) and limit for $\alpha \to \infty$ according to Eq. (\ref{Qcreq}) (green),
	 		 (b) $\Phi(r_H)(\alpha)$ (red) and relation from Eq. (\ref{Phieq}) (green),
	 		 (c) $\Phi'(r_H)(\alpha)$ (red) and expansion coefficient $\Phi_1(r_H)(\alpha)$ from Eq. (\ref{phi1}) (green),
	 		 (d) $m'(r_H)(\alpha)$ (red) and expansion coefficient $m_1(r_H)(\alpha)$ from Eq. (\ref{phi1}) (green).
}
\label{fig5}
\end{figure}

The figure shows that with decreasing $\alpha$, the critical radius $r_{\rm cr}$ increases.
We recall that it coincides with the critical charge, $r_{\rm cr}=Q_{\rm cr}$.
At the same time, the scalar field assumes larger values in the interior.
We highlight the $\alpha$-dependence of the critical charge $Q_{\rm cr}$ and 
of the horizon value of the scalar field $\Phi(r_H)$ in Fig.~\ref{fig5}(a)
and Fig.~\ref{fig5}(b), respectively.
We note the steep increase of the critical charge $Q_{\rm cr}$ for small $\alpha$,
while $Q_{\rm cr} \to r_H$ for large $\alpha$.
This dependence can be well described by the simple relation
\begin{equation}
\frac{Q_{\rm cr}}{r_H}-1=\frac{1}{4\sqrt{2}\, \alpha} \ , %\frac{r_{H}^{4}}{64\sqrt{2}\, \alpha}
\label{Qcreq}
\end{equation}
as demonstrated in the figure.
The horizon value of the scalar field $\Phi(r_H)$ satisfies the even simpler relation
\begin{equation}
\Phi(r_H) =\frac{1}{\sqrt{\alpha}} \ , %\frac{1}{\sqrt{\alpha}}
\label{Phieq}
\end{equation}
as seen in the figure as well.
In Fig.~\ref{fig5}(c) and Fig.~\ref{fig5}(d), we show that for the interior critical solution,
the derivatives of the metric function $m(r)$ and the scalar function $\Phi(r)$ at the horizon
precisely respect the expansion at the horizon, Eqs.~(\ref{nhe}) and (\ref{nhe2}), respectively.

\section{Excited solutions}

\begin{figure}[h!]
	\centering
	\includegraphics[width=0.38\linewidth,angle=-90]{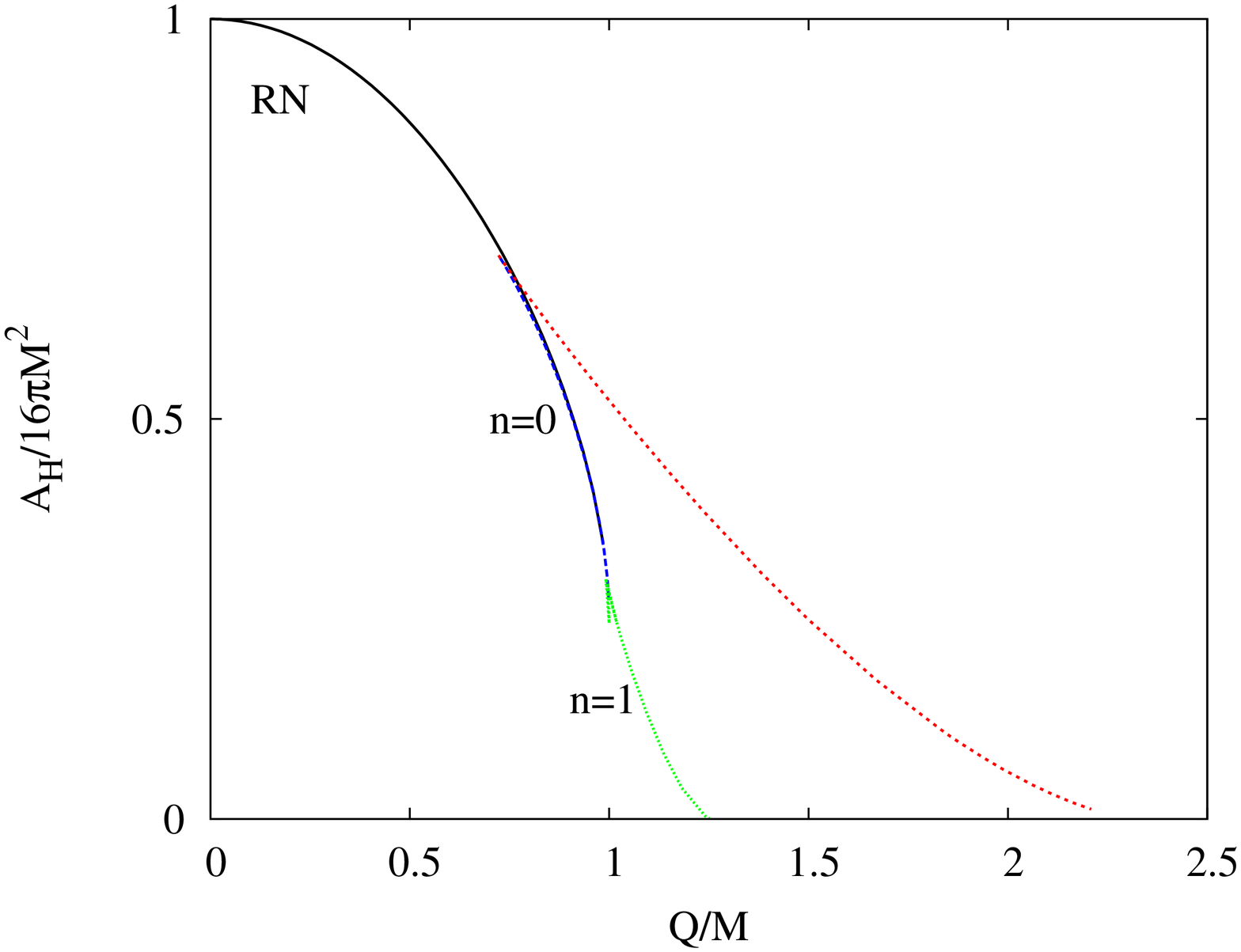}
	\includegraphics[width=0.38\linewidth,angle=-90]{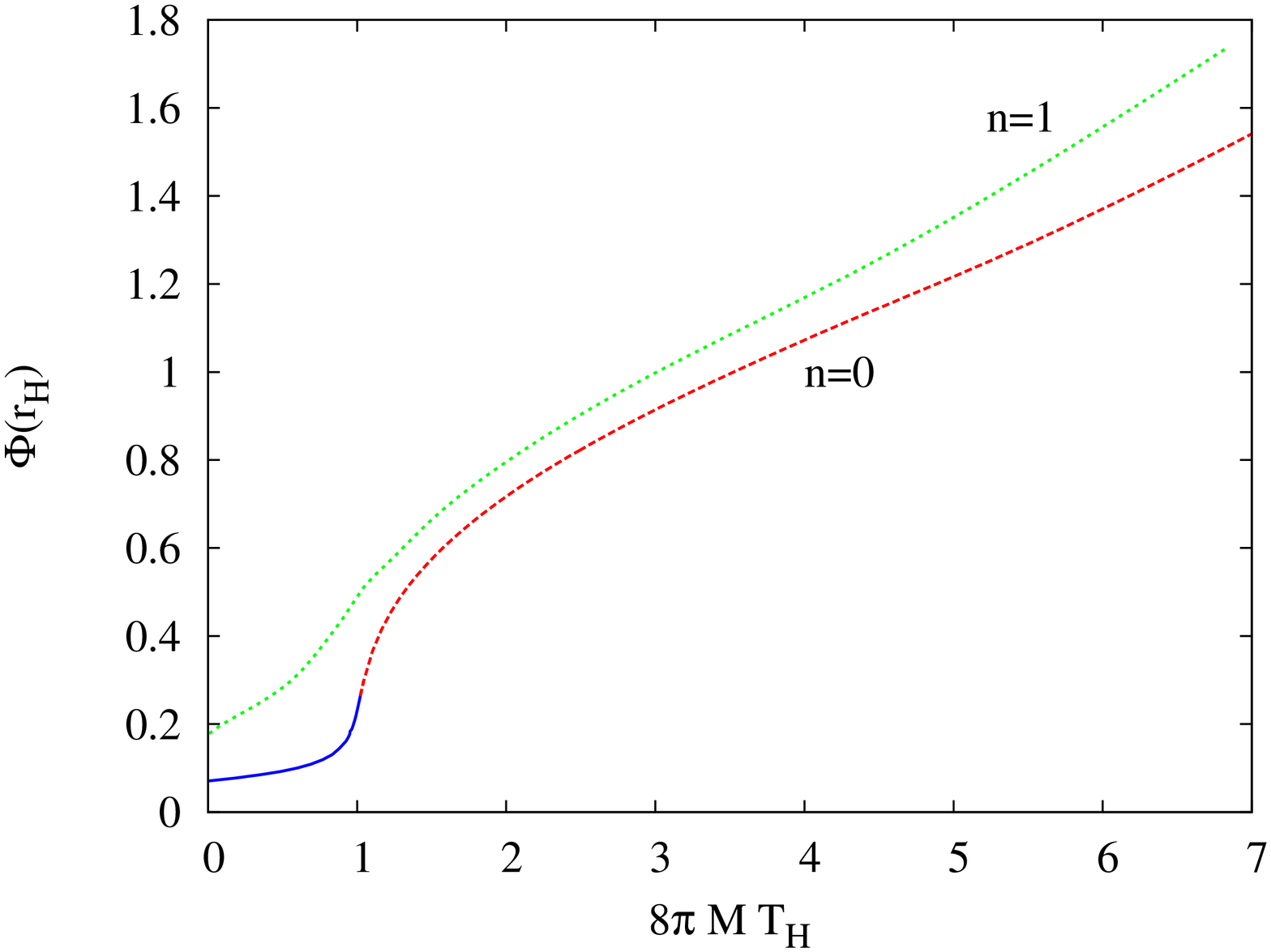}
	\caption{
		(a) Reduced horizon area $a_H$ vs. $q$ for $\alpha=200$. In addition to the RN branch (black), the ($n=0$) cold branch (blue) and the ($n=0$) hot branch (red), the $n=1$ excited solutions (green) are shown.
		(b) Similar figure for the scalar field at the horizon $\Phi_H$ vs. reduced temperature $t_H$.
	}
	\label{fig_n1}
\end{figure}

Until now we have focused the discussion on the fundamental branch of scalarized black holes, with scalar field functions that possess no nodes. However, in certain cases, the model allows for the existence of excited solutions, meaning solutions with nodes in the scalar field functions. Let us discuss them briefly here.

These solutions present very similar properties to the $n=0$ solutions, with a hot/cold branch structure, the hot one bifurcating from the cold one. The cold branch also reaches a critical end-point with $q=1$. However, an important difference is that all of these excited branches are radially unstable, as previously discussed in \cite{Blazquez-Salcedo:2020nhs,Blazquez-Salcedo:2020jee}. Also, the number of excited branches depends on the coupling constant $\alpha$, i.e. the larger the value of $\alpha$, the more excited branches exist. 

As an example, in Fig. \ref{fig_n1}(a) we show again the reduced area $a_H$ as a function of the reduced charge $q$ for $\alpha=200$. This is the same as Fig. \ref{fig1}(b), but now we include the $n=1$ solutions in green. As we can see, these excited solutions could also be distinguished in two different branches: One branch (cold) would extend from $q=1$ to a minimum $q_{min}<1$; the second branch (hot) would extend from this $q_{min}$ up to a certain $q_{max}>1$. It turns out that the interval $(q_{min},q_{max})$, where the $n=1$ solutions exist, is smaller than the domain interval of the $n=0$ solutions. On the other hand, for this particular value of the coupling constant ($\alpha=200$), only $n=1$ excited solutions exist. However, sufficiently large values of $\alpha$ allow for additional branches with $n>1$ excited solutions to appear.

In Fig. \ref{fig_n1}(b), we show the scalar field at the horizon as a function of the reduced temperature $t_H$, also for $\alpha=200$. As we approach the $q=1$ limit along the cold branch, the temperature goes to zero, but the value of the scalar field remains constant. (This is similar to what happens for the $n=0$ solutions we have discussed in the previous sections, as shown in this figure). If we compare solutions with the same $t_h$, the larger the excitation number, the larger is the value of the scalar field at the horizon. 

As already said, the critical behavior is also present in these excited solutions. When fixing $r_H=2$, the critical solutions possess a value of the electric charge $Q=Q_{cr}(n)>2$ (this value depends on $n$). 
The limit results in a set of critical solutions with split spacetime, similar to the $n=0$ solutions we have been discussing in the previous sections. The exterior part is the extremal RN solution. On the interior part, however, we have a non-trivial scalar field whose number of nodes can be labeled by an integer number $n$.

\begin{figure}[h!]
	\centering
	\includegraphics[width=0.38\linewidth,angle=-90]{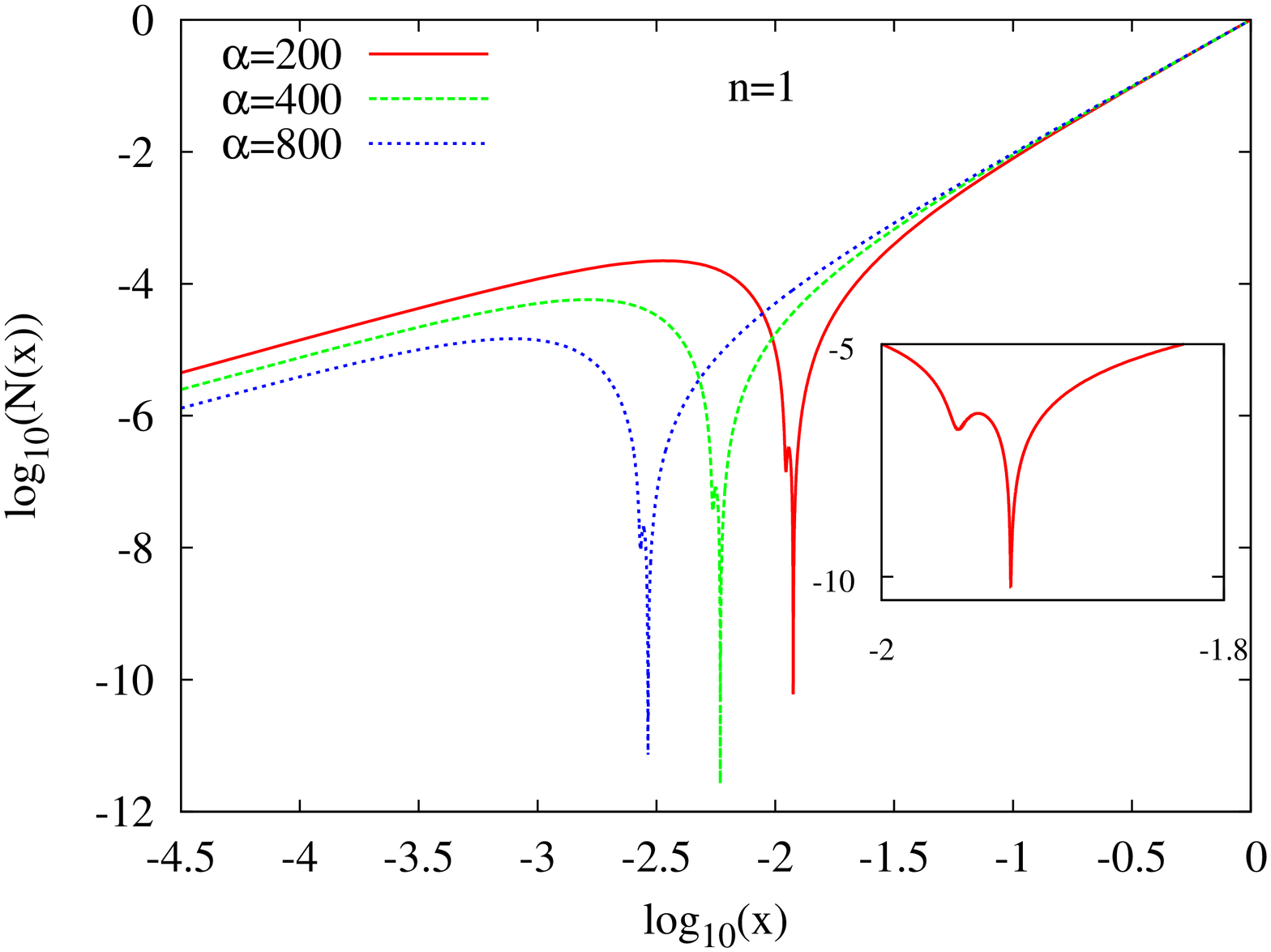}
	\includegraphics[width=0.38\linewidth,angle=-90]{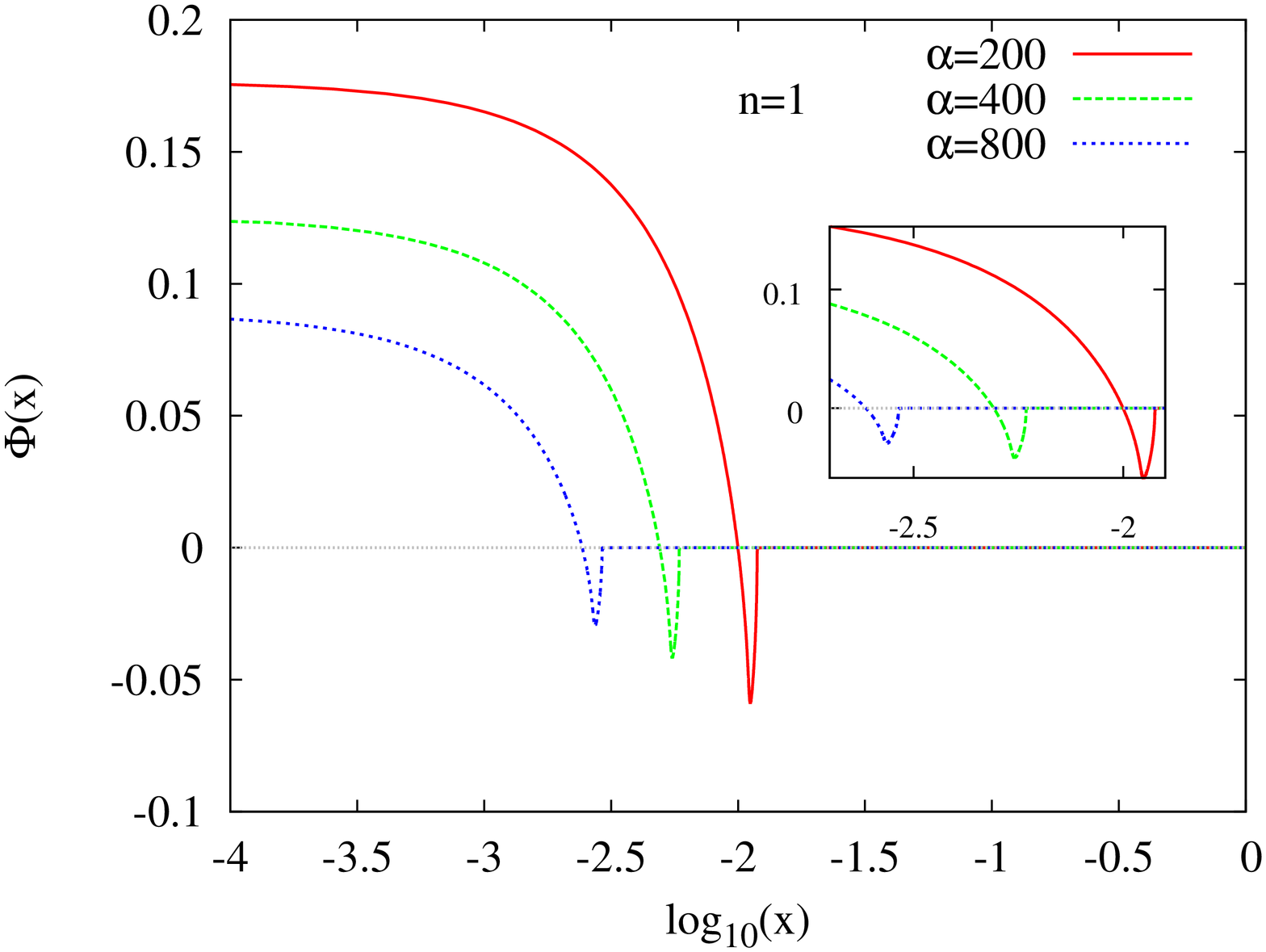}
	\caption{
	 	Critical solutions with $n=1$ for a set of couplings $\alpha$:
(a)  metric function $N(r)=1-\frac{2m(r)}{r}$,
(b)  scalar function $\Phi(r)$
vs. the compactified radial coordinate $x=1-\frac{r_H}{r}$. 
	}
	\label{fig_profiles_n1}
\end{figure}

In Fig. \ref{fig_profiles_n1}, some profiles for $n=1$ critical solutions with different values of $\alpha$ are shown as an example. In particular, we show the critical solutions for $\alpha=200$ with $Q_{cr}=2.02407591$ (red), for $\alpha=400$ with $Q_{cr}=2.01180482$ (green), and for $\alpha=800$ with $Q_{cr}=2.00584391$ (blue).

In Fig. \ref{fig_profiles_n1}(a), we show the metric function $N(r)$ versus $x=1-r_H/r$. It demonstrates that the behavior of these $n=1$ solutions is very similar to the one presented in Fig. \ref{fig4}(a), which corresponds to the $n=0$ solutions. Essentially, the metric function $N(r)$ develops a zero at some critical value $r=r_{cr}$. For $r>r_{cr}$, the solution is extremal RN, but in the interior region it is a non-trivial solution with a scalar field. The main difference now, as shown more clearly in the inset, is that the function develops a minimum at a certain point $r<r_{cr}$.

Fig. \ref{fig_profiles_n1}(b) exhibits the scalar field $\Phi$ versus $x$, with the inset showing a zoomed region around the critical value of the radial coordinate $r_{cr}$. This demonstrates that the profile of the scalar field is quite different from the one in Fig.  \ref{fig4}(b). Now the scalar field function not only vanishes at $r_{cr}$, it also has a node at some intermediate point.

\begin{figure}[h!]
	\centering
	\includegraphics[width=0.38\linewidth,angle=-90]{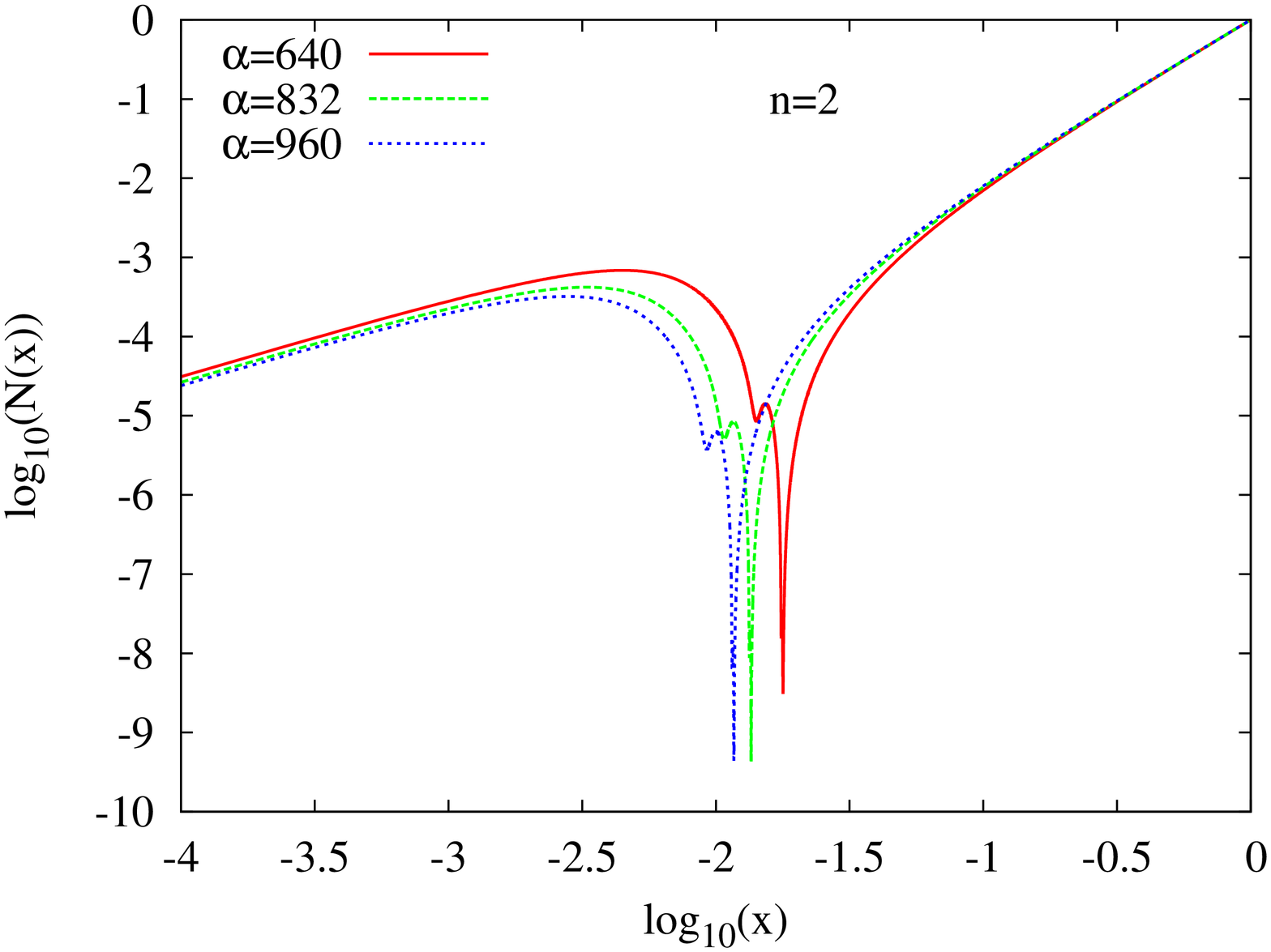}
	\includegraphics[width=0.38\linewidth,angle=-90]{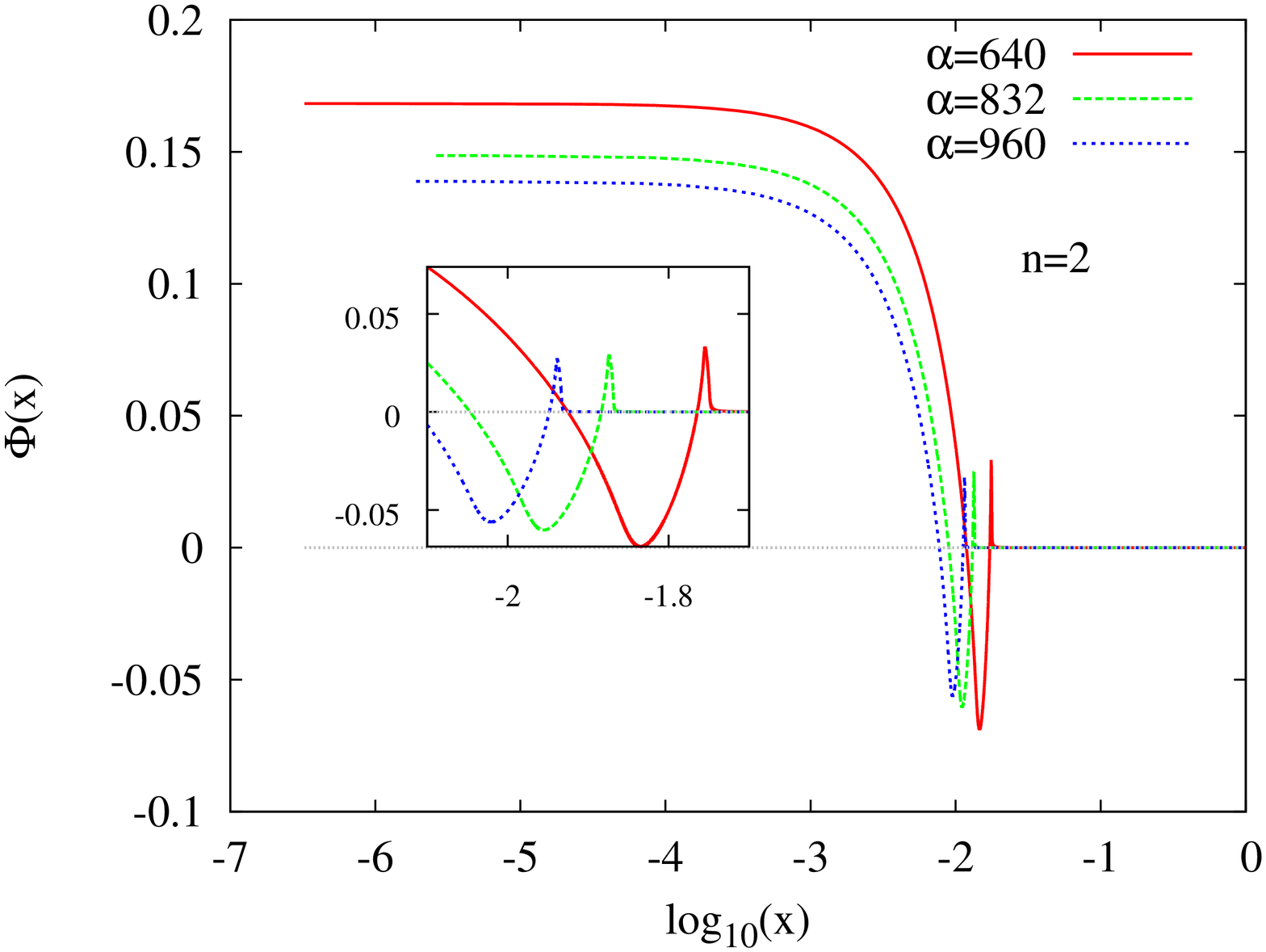}

	\caption{
	 	Critical solutions with $n=2$ for a set of couplings $\alpha$:
(a)  metric function $N(r)=1-\frac{2m(r)}{r}$,
(b)  scalar function $\Phi(r)$
vs. the compactified radial coordinate $x=1-\frac{r_H}{r}$. 
	}
	\label{fig_profiles_n2}
\end{figure}

Solutions with more nodes also follow this pattern. As an example, we depict in Fig. \ref{fig_profiles_n2} similar plots for $n=2$ solutions. These correspond to critical solutions for $\alpha=640$ with $Q_{cr}=2.0362514$ (red), for $\alpha=832$ with $Q_{cr}=2.0274443$ (green), and for $\alpha=960$ with $Q_{cr}=2.02361315$ (blue). Fig. \ref{fig_profiles_n2}(left) exhibits how the $N$ function develops a minimum a bit below $r_{cr}$; and in Fig. \ref{fig_profiles_n2}(right), we demonstrate that the scalar field function has two nodes below $r_{cr}$.

\section{Conclusion}

We have considered black holes in EMs models with a quartic coupling function
$f(\Phi) = 1 + \alpha \Phi^4$,
that allows for RN black holes, as well as cold and hot scalarized black holes
\cite{Blazquez-Salcedo:2020nhs}.
The models are parameterized by the coupling constant $\alpha$
and exhibit generic features. In particular, the RN black holes
never become unstable to grow scalar hair
\cite{Blazquez-Salcedo:2020nhs,Blazquez-Salcedo:2020jee}.
Still, the cold branch follows the RN branch to a large extent,
and starts at $q=Q/M=1$, the endpoint of the RN branch.

In particular, we have investigated the approach of the cold branch
to this critical point $q=1$ in detail for these EMs models.
For that purpose, we have fixed the horizon radius $r_H$ of the black holes
and then increased the electromagnetic charge $Q$
until for some fixed coupling $\alpha$, a critical solution with $q=1$ has been reached.
Whereas an extremal RN black hole with horizon radius $r_H$ satisfies
$q=1$ with $r_H=Q=M$, the critical solution on the cold branch satisfies
$q=1$ with $r_{\rm cr}=Q_{\rm cr}=M_{\rm cr}$ and $r_H<r_{\rm cr}$.
Consequently, the critical solution cannot simply be an extremal RN black hole.

Inspection of the functions of the critical solution then revealed
that the critical solution splits the spacetime into two parts: an exterior part which indeed corresponds to an extremal RN black hole
solution, but however with $r_{\rm cr}=Q_{\rm cr}=M_{\rm cr}$,
and an interior part, which has a finite scalar field that vanishes at $r_{\rm cr}$
and a vanishing electromagnetic field.
Since the radial metric function develops a double zero at $r_{\rm cr}$,
both parts of the spacetime become infinite in size as the spacetime splits.

By studying the critical  solution for many values of the coupling $\alpha$,
we have shown that this observed phenomenon is generic for these EMs models.
In fact, the critical charge $Q_{\rm cr}$ possesses a rather simple $\alpha$-dependence, with $(Q_{\rm cr}/r_H-1)$ being inversely proportional to $\alpha$,
while the horizon value of the scalar field $\Phi_H$ 
is inversely proportional the square root of $\alpha$.
At the same time, the functions of the critical solution satisfy the
horizon expansion at $r_H$, as well as the expansion at infinity, the latter
of course with vanishing scalar charge.

In addition, we have discussed the case of the excited solutions, with nodes in the scalar field functions. These solutions posses a similar (hot/cold) branch structure, with a critical limit that splits the spacetime just like in the nodeless case. The excited solutions can be labeled by an integer number $n$, counting the number of nodes of the scalar in the interior part of the spacetime.

It is interesting that the critical phenomenon which has previously been encountered
for non-Abelian magnetic monopoles and their associated hairy black holes,
as well as for similar non-Abelian solutions,
arises also in these EMs models.
Note a somewhat similar observation in the recent investigation of
strong gravity effects of charged $Q$-clouds and inflating black holes
\cite{Brihaye:2020vce} (see also \cite{Brihaye:2020yuv} in this issue).
For non-Abelian monopoles, the interior solution has non-trivial non-Abelian
gauge fields and Higgs field, whereas the exterior solution is simply an 
embedded RN black hole with magnetic charge. Not surprisingly, 
for solutions with both electric and magnetic charge,
this phenomenon persists.
Even pure non-Abelian solutions (without a Higgs field) were shown
to exhibit a somewhat analogous phenomenon, which emerges in the
limit of infinite node number (see e.g.~\cite{Volkov:1998cc}).
It would thus be interesting to find general criteria to identify models
that allow for this type of phenomenon where the spacetime splits
into two infinitely extended parts, with the interior part containing
non-trivial fields that vanish identically in the exterior part,
where a simple extremal black hole solution emerges.

%I put only extremal black hole and not extremal RN because the phenomenon is also present for non-Abelian EYMd solutions that tend to an Abelian EMd black hole in the exterior in the limit. We could mention it, but I refrained from it.

\paragraph{Acknowledgments}
JLBS, SK and JK gratefully acknowledge support by the
DFG Research Training Group 1620  \textit{Models of Gravity}.
%and the COST Action CA16104 \textit{GWverse}.
JLBS would like to acknowledge support from
the DFG Project No. BL 1553. 
{We also would like to acknowledge networking support by the COST Actions CA15117 and CA16104.}
\appendix

\bibliographystyle{ieeetr}
\bibliography{notes}

\begin{thebibliography}{10}

\bibitem{Doneva:2017bvd}
D.~D. Doneva and S.~S. Yazadjiev, ``{New Gauss-Bonnet Black Holes with
  Curvature-Induced Scalarization in Extended Scalar-Tensor Theories},'' {\em
  Phys. Rev. Lett.}, vol.~120, no.~13, p.~131103, 2018.

\bibitem{Antoniou:2017acq}
G.~Antoniou, A.~Bakopoulos, and P.~Kanti, ``{Evasion of No-Hair Theorems and
  Novel Black-Hole Solutions in Gauss-Bonnet Theories},'' {\em Phys. Rev.
  Lett.}, vol.~120, no.~13, p.~131102, 2018.

\bibitem{Silva:2017uqg}
H.~O. Silva, J.~Sakstein, L.~Gualtieri, T.~P. Sotiriou, and E.~Berti,
  ``{Spontaneous scalarization of black holes and compact stars from a
  Gauss-Bonnet coupling},'' {\em Phys. Rev. Lett.}, vol.~120, no.~13,
  p.~131104, 2018.

\bibitem{Antoniou:2017hxj}
G.~Antoniou, A.~Bakopoulos, and P.~Kanti, ``{Black-Hole Solutions with Scalar
  Hair in Einstein-Scalar-Gauss-Bonnet Theories},'' {\em Phys. Rev. D},
  vol.~97, no.~8, p.~084037, 2018.

\bibitem{Blazquez-Salcedo:2018jnn}
J.~L. Blázquez-Salcedo, D.~D. Doneva, J.~Kunz, and S.~S. Yazadjiev, ``{Radial
  perturbations of the scalarized Einstein-Gauss-Bonnet black holes},'' {\em
  Phys. Rev. D}, vol.~98, no.~8, p.~084011, 2018.

\bibitem{Doneva:2018rou}
D.~D. Doneva, S.~Kiorpelidi, P.~G. Nedkova, E.~Papantonopoulos, and S.~S.
  Yazadjiev, ``{Charged Gauss-Bonnet black holes with curvature induced
  scalarization in the extended scalar-tensor theories},'' {\em Phys. Rev. D},
  vol.~98, no.~10, p.~104056, 2018.

\bibitem{Minamitsuji:2018xde}
M.~Minamitsuji and T.~Ikeda, ``{Scalarized black holes in the presence of the
  coupling to Gauss-Bonnet gravity},'' {\em Phys. Rev. D}, vol.~99, no.~4,
  p.~044017, 2019.

\bibitem{Silva:2018qhn}
H.~O. Silva, C.~F. Macedo, T.~P. Sotiriou, L.~Gualtieri, J.~Sakstein, and
  E.~Berti, ``{Stability of scalarized black hole solutions in
  scalar-Gauss-Bonnet gravity},'' {\em Phys. Rev. D}, vol.~99, no.~6,
  p.~064011, 2019.

\bibitem{Brihaye:2018grv}
Y.~Brihaye and L.~Ducobu, ``{Hairy black holes, boson stars and non-minimal
  coupling to curvature invariants},'' {\em Phys. Lett. B}, vol.~795,
  pp.~135--143, 2019.

\bibitem{Doneva:2019vuh}
D.~D. Doneva, K.~V. Staykov, and S.~S. Yazadjiev, ``{Gauss-Bonnet black holes
  with a massive scalar field},'' {\em Phys. Rev. D}, vol.~99, no.~10,
  p.~104045, 2019.

\bibitem{Myung:2019wvb}
Y.~S. Myung and D.-C. Zou, ``{Black holes in Gauss--Bonnet and
  Chern--Simons-scalar theory},'' {\em Int. J. Mod. Phys. D}, vol.~28, no.~09,
  p.~1950114, 2019.

\bibitem{Cunha:2019dwb}
P.~V. Cunha, C.~A. Herdeiro, and E.~Radu, ``{Spontaneously Scalarized Kerr
  Black Holes in Extended Scalar-Tensor--Gauss-Bonnet Gravity},'' {\em Phys.
  Rev. Lett.}, vol.~123, no.~1, p.~011101, 2019.

\bibitem{Macedo:2019sem}
C.~F. Macedo, J.~Sakstein, E.~Berti, L.~Gualtieri, H.~O. Silva, and T.~P.
  Sotiriou, ``{Self-interactions and Spontaneous Black Hole Scalarization},''
  {\em Phys. Rev. D}, vol.~99, no.~10, p.~104041, 2019.

\bibitem{Hod:2019pmb}
S.~Hod, ``{Spontaneous scalarization of Gauss-Bonnet black holes: Analytic
  treatment in the linearized regime},'' {\em Phys. Rev. D}, vol.~100, no.~6,
  p.~064039, 2019.

\bibitem{Collodel:2019kkx}
L.~G. Collodel, B.~Kleihaus, J.~Kunz, and E.~Berti, ``{Spinning and excited
  black holes in Einstein-scalar-Gauss--Bonnet theory},'' {\em Class. Quant.
  Grav.}, vol.~37, no.~7, p.~075018, 2020.

\bibitem{Bakopoulos:2020dfg}
A.~Bakopoulos, P.~Kanti, and N.~Pappas, ``{Large and ultracompact Gauss-Bonnet
  black holes with a self-interacting scalar field},'' {\em Phys. Rev. D},
  vol.~101, no.~8, p.~084059, 2020.

\bibitem{Blazquez-Salcedo:2020rhf}
J.~L. Blázquez-Salcedo, D.~D. Doneva, S.~Kahlen, J.~Kunz, P.~Nedkova, and
  S.~S. Yazadjiev, ``{Axial perturbations of the scalarized
  Einstein-Gauss-Bonnet black holes},'' {\em Phys. Rev. D}, vol.~101, no.~10,
  p.~104006, 2020.

\bibitem{Blazquez-Salcedo:2020caw}
J.~L. Bl\'azquez-Salcedo, D.~D. Doneva, S.~Kahlen, J.~Kunz, P.~Nedkova, and
  S.~S. Yazadjiev, ``{Polar quasinormal modes of the scalarized
  Einstein-Gauss-Bonnet black holes},'' {\em arXiv}, 2006.06006, 6 2020.

\bibitem{Dima:2020yac}
A.~Dima, E.~Barausse, N.~Franchini, and T.~P. Sotiriou, ``{Spin-induced black
  hole spontaneous scalarization},'' {\em arXiv}, 2006.03095, 6 2020.

\bibitem{Doneva:2020nbb}
D.~D. Doneva, L.~G. Collodel, C.~J. Kr\"uger, and S.~S. Yazadjiev, ``{Black
  hole scalarization induced by the spin -- 2+1 time evolution},'' {\em arXiv}, 2008.07391, 8 2020.

\bibitem{Berti:2020kgk}
E.~Berti, L.~G. Collodel, B.~Kleihaus, and J.~Kunz, ``{Spin-induced black-hole
  scalarization in Einstein-scalar-Gauss-Bonnet theory},'' {\em arXiv}, 2009.03905, 9 2020.

\bibitem{Herdeiro:2020wei}
C.~A. Herdeiro, E.~Radu, H.~O. Silva, T.~P. Sotiriou, and N.~Yunes,
  ``{Spin-induced scalarized black holes},'' {\em arXiv}, 2009.03904, 9 2020.

\bibitem{Herdeiro:2018wub}
C.~A. Herdeiro, E.~Radu, N.~Sanchis-Gual, and J.~A. Font, ``{Spontaneous
  Scalarization of Charged Black Holes},'' {\em Phys. Rev. Lett.}, vol.~121,
  no.~10, p.~101102, 2018.

\bibitem{Myung:2018vug}
Y.~S. Myung and D.-C. Zou, ``{Instability of Reissner--Nordström black hole in
  Einstein-Maxwell-scalar theory},'' {\em Eur. Phys. J. C}, vol.~79, no.~3,
  p.~273, 2019.

\bibitem{Boskovic:2018lkj}
M.~Boskovic, R.~Brito, V.~Cardoso, T.~Ikeda, and H.~Witek, ``{Axionic
  instabilities and new black hole solutions},'' {\em Phys. Rev. D}, vol.~99,
  no.~3, p.~035006, 2019.

\bibitem{Myung:2018jvi}
Y.~S. Myung and D.-C. Zou, ``{Quasinormal modes of scalarized black holes in
  the Einstein--Maxwell--Scalar theory},'' {\em Phys. Lett. B}, vol.~790,
  pp.~400--407, 2019.

\bibitem{Fernandes:2019rez}
P.~G. Fernandes, C.~A. Herdeiro, A.~M. Pombo, E.~Radu, and N.~Sanchis-Gual,
  ``{Spontaneous Scalarisation of Charged Black Holes: Coupling Dependence and
  Dynamical Features},'' {\em Class. Quant. Grav.}, vol.~36, no.~13, p.~134002,
  2019.
\newblock [Erratum: Class.Quant.Grav. 37, 049501 (2020)].

\bibitem{Brihaye:2019kvj}
Y.~Brihaye and B.~Hartmann, ``{Spontaneous scalarization of charged black holes
  at the approach to extremality},'' {\em Phys. Lett. B}, vol.~792,
  pp.~244--250, 2019.

\bibitem{Herdeiro:2019oqp}
C.~A. Herdeiro and J.~M. Oliveira, ``{On the inexistence of solitons in
  Einstein--Maxwell-scalar models},'' {\em Class. Quant. Grav.}, vol.~36,
  no.~10, p.~105015, 2019.

\bibitem{Myung:2019oua}
Y.~S. Myung and D.-C. Zou, ``{Stability of scalarized charged black holes in
  the Einstein--Maxwell--Scalar theory},'' {\em Eur. Phys. J. C}, vol.~79,
  no.~8, p.~641, 2019.

\bibitem{Astefanesei:2019pfq}
D.~Astefanesei, C.~Herdeiro, A.~Pombo, and E.~Radu, ``{Einstein-Maxwell-scalar
  black holes: classes of solutions, dyons and extremality},'' {\em JHEP},
  vol.~10, p.~078, 2019.

\bibitem{Konoplya:2019goy}
R.~Konoplya and A.~Zhidenko, ``{Analytical representation for metrics of
  scalarized Einstein-Maxwell black holes and their shadows},'' {\em Phys. Rev.
  D}, vol.~100, no.~4, p.~044015, 2019.

\bibitem{Fernandes:2019kmh}
P.~G. Fernandes, C.~A. Herdeiro, A.~M. Pombo, E.~Radu, and N.~Sanchis-Gual,
  ``{Charged black holes with axionic-type couplings: Classes of solutions and
  dynamical scalarization},'' {\em Phys. Rev. D}, vol.~100, no.~8, p.~084045,
  2019.

\bibitem{Herdeiro:2019tmb}
C.~A. Herdeiro and J.~M. Oliveira, ``{On the inexistence of self-gravitating
  solitons in generalised axion electrodynamics},'' {\em Phys. Lett. B},
  vol.~800, p.~135076, 2020.

\bibitem{Zou:2019bpt}
D.-C. Zou and Y.~S. Myung, ``{Scalarized charged black holes with scalar mass
  term},'' {\em Phys. Rev. D}, vol.~100, no.~12, p.~124055, 2019.

\bibitem{Brihaye:2019gla}
Y.~Brihaye, C.~Herdeiro, and E.~Radu, ``{Black Hole Spontaneous Scalarisation
  with a Positive Cosmological Constant},'' {\em Phys. Lett. B}, vol.~802,
  p.~135269, 2020.

\bibitem{Astefanesei:2019qsg}
D.~Astefanesei, J.~L. Bl\'azquez-Salcedo, C.~Herdeiro, E.~Radu, and
  N.~Sanchis-Gual, ``{Dynamically and thermodynamically stable black holes in
  Einstein-Maxwell-dilaton gravity},'' {\em JHEP}, vol.~07, p.~063, 2020.

\bibitem{Blazquez-Salcedo:2020nhs}
J.~L. Bl\'azquez-Salcedo, C.~A. Herdeiro, J.~Kunz, A.~M. Pombo, and E.~Radu,
  ``{Einstein-Maxwell-scalar black holes: the hot, the cold and the bald},''
  {\em Phys. Lett. B}, vol.~806, p.~135493, 2020.

\bibitem{Blazquez-Salcedo:2020jee}
J.~L. Bl\'azquez-Salcedo, C.~A. Herdeiro, S.~Kahlen, J.~Kunz, A.~M. Pombo, and
  E.~Radu, ``{Quasinormal modes of hot, cold and bald Einstein-Maxwell-scalar
  black holes},'' {\em arXiv}, 2008.11744, 8 2020.

\bibitem{Astefanesei:2020xvn}
D.~Astefanesei, J.~L. Bl\'azquez-Salcedo, F.~G\'omez, and R.~Rojas,
  ``{Thermodynamically stable asymptotically flat hairy black holes with a
  dilaton potential: the general case},'' {\em arXiv}, 2009.01854, 9 2020.

\bibitem{Gibbons:1987ps}
G.~Gibbons and K.-i. Maeda, ``{Black Holes and Membranes in Higher Dimensional
  Theories with Dilaton Fields},'' {\em Nucl. Phys. B}, vol.~298, pp.~741--775,
  1988.

\bibitem{Kanti:1995vq}
P.~Kanti, N.~Mavromatos, J.~Rizos, K.~Tamvakis, and E.~Winstanley, ``{Dilatonic
  black holes in higher curvature string gravity},'' {\em Phys. Rev. D},
  vol.~54, pp.~5049--5058, 1996.

\bibitem{Volkov:1998cc}
M.~S. Volkov and D.~V. Gal'tsov, ``{Gravitating non-Abelian solitons and black
  holes with Yang-Mills fields},'' {\em Phys. Rept.}, vol.~319, pp.~1--83,
  1999.

\bibitem{Galtsov:2001myk}
D.~Gal'tsov, ``{Gravitating lumps},'' in {\em {16th International Conference on
  General Relativity and Gravitation (GR16)}}, pp.~142--161, 2013.

\bibitem{Kleihaus:2016rgf}
B.~Kleihaus, J.~Kunz, and F.~Navarro-Lerida, ``{Rotating black holes with
  non-Abelian hair},'' {\em Class. Quant. Grav.}, vol.~33, no.~23, p.~234002,
  2016.

\bibitem{Lee:1991vy}
K.-M. Lee, V.~Nair, and E.~J. Weinberg, ``{Black holes in magnetic
  monopoles},'' {\em Phys. Rev. D}, vol.~45, pp.~2751--2761, 1992.

\bibitem{Breitenlohner:1991aa}
P.~Breitenlohner, P.~Forgacs, and D.~Maison, ``{Gravitating monopole
  solutions},'' {\em Nucl. Phys. B}, vol.~383, pp.~357--376, 1992.

\bibitem{Breitenlohner:1994di}
P.~Breitenlohner, P.~Forgacs, and D.~Maison, ``{Gravitating monopole solutions.
  2},'' {\em Nucl. Phys. B}, vol.~442, pp.~126--156, 1995.

\bibitem{Ridgway:1994sm}
S.~Ridgway and E.~J. Weinberg, ``{Instabilities of magnetically charged black
  holes},'' {\em Phys. Rev. D}, vol.~51, pp.~638--646, 1995.

\bibitem{Brihaye:1998cm}
Y.~Brihaye, B.~Hartmann, and J.~Kunz, ``{Gravitating dyons and dyonic black
  holes},'' {\em Phys. Lett. B}, vol.~441, pp.~77--82, 1998.

\bibitem{Hartmann:2000gx}
B.~Hartmann, B.~Kleihaus, and J.~Kunz, ``{Gravitationally bound monopoles},''
  {\em Phys. Rev. Lett.}, vol.~86, pp.~1422--1425, 2001.

\bibitem{Hartmann:2001ic}
B.~Hartmann, B.~Kleihaus, and J.~Kunz, ``{Axially symmetric monopoles and black
  holes in Einstein-Yang-Mills-Higgs theory},'' {\em Phys. Rev. D}, vol.~65,
  p.~024027, 2002.

\bibitem{Ascher:1979iha}
U.~Ascher, J.~Christiansen, and R.~Russell, ``{A Collocation Solver for Mixed
  Order Systems of Boundary Value Problems},'' {\em Math. Comput.}, vol.~33,
  no.~146, pp.~659--679, 1979.

\bibitem{Brihaye:2020vce}
Y.~Brihaye and B.~Hartmann, ``{Strong gravity effects of charged Q-clouds and
  inflating black holes},'' {\em arXiv}, 2009.08293, 9 2020.

\bibitem{Brihaye:2020yuv}
Y.~Brihaye, F.~C\^onsole, and B.~Hartmann, ``{Inflation inside non-topological
  defects and scalar black holes},'' {\em arXiv}, 2010.15625, 10 2020.

\end{thebibliography}

\end{document}